\documentclass[letterpaper,12pt,peerreviewca,draftcls]{IEEEtran}

\usepackage[margin=1in]{geometry}
\usepackage[compress]{cite}
\usepackage{url}

\usepackage{times}
\usepackage{graphicx}        
\usepackage{multicol}        
\usepackage[bottom]{footmisc}
\usepackage{amssymb,amsmath,amsthm,mathrsfs}
\usepackage{mathtools}
\usepackage{enumerate}
\usepackage{enumitem}
\usepackage{framed}
\usepackage{algorithm,algorithmic,psfrag}
\usepackage{soul}
\usepackage{color}
\usepackage[colorlinks=true, pdfstartview=FitV, linkcolor=black, citecolor=black, urlcolor=blue]{hyperref}
\usepackage[dvipsnames]{xcolor}
\usepackage{bm}
\usepackage{bbm}

\definecolor{coolBlue}{rgb}{.3, .5, 1}
\definecolor{rosePink}{rgb}{.9, .5, .4}
\definecolor{ghost}{rgb}{.8, .8, .8}
\definecolor{paleGreen}{rgb}{.3, .7, .3}

\def\BibTeX{{\rm B\kern-.05em{\sc i\kern-.025em b}\kern-.08em
    T\kern-.1667em\lower.7ex\hbox{E}\kern-.125emX}}

\title{The Milieu, Science \&\ Logic of Feedback Control} 
\author{Robert R. Bitmead\\
Department of Mechanical \&\ Aerospace Engineering, University of California San Diego\\
rbitmead@ucsd.edu, \today}
\begin{document}
\maketitle

\section*{Roadmap}

\textbf{Sections and Sidebars}
\begin{enumerate}[label=\arabic*.]
\item Section: Introduction
\item Section: Intentions of control
\begin{enumerate}[left=5mm,label=\theenumi\arabic*.]
\item {Sidebar: Agreeing to terms}
\end{enumerate}
\item Section: The epistemology of control
\item Section: The philosophies of science and control
\item Section: The logic of control
\begin{enumerate}[left=5mm,label=\theenumi\arabic*.]
\item {Sidebar: Control theory is deductive}
\item {Sidebar: Induction and (purely) data-driven methods}
\end{enumerate}
\item Section: Models and model-based control
\item Section: Observables, unobservables and states
\item Section: Industrial examples and the role of data in control
\begin{enumerate}[left=5mm,label=\theenumi\arabic*.]
\item {Sidebar: DC servomotor identification and control}
\item {Sidebar: FrothSense+ an inductively trained instrument}
\item {Sidebar: Combustion Instability modeling and control}
\item {Sidebar: Control of a sugar mill}
\item {Sidebar: Countermanding experience}
\end{enumerate}
\item Section: Joining the threads
\item Acknowledgements
\item References
\end{enumerate}

\section{Introduction}\label{section:Introduction}
\begin{quote}
\textit{``The cardinal sin in control is to believe that the plant is given''} Karl \AA str\"om \cite{AstromCardinal:2003}.
\end{quote}
\AA str\"om, a towering figure of control theory and practice and awardee of the 1993 IEEE Medal of Honor for his work on adaptive control, provides this assessment of the obstinate part of realizing a feedback controller. And yet we are exhorted to rely on solely-data-driven methods of control design skipping the modeling and plant identification phases entirely. What is going on? Whom should we trust? How do we reconcile the implied ease (or indeed avoidance) of modeling with the steely focus on robustness of the control and the capacity of feedback to accommodate uncertainty? This paper seeks to investigate this subject with the objective of appreciating not whom to trust but what are the circumstances where the direct paradigm of control design from any lightly qualified data set provides a sensible way forward. Here is a clue: It depends on the confidence of your contentions about the plant system, the detailed data themselves and your appetite for failure.

There are constant reminders that current times are data-drenched and purportedly transformative. This paper is written with the intention of coming to grips with the fact that \textit{data} is a noun and that it needs to be preceded by adjectives describing its suitability, or unsuitability, for purpose and not just its cardinality. Informative data also comes with a cost and, in many circumstances, is both rare and grueling to generate. Experiments consume time and resources which costs need to be balanced against potential benefits. There are evidently problems where copious ambient data suffice for decision-making at an adequate level of accuracy; but it is not a universal solution or the failure rate might be unacceptably high. There are many contemporary thinkers who have entered the fray to temper the hype and make sense of the human (or controller) in the loop -- a \textit{guide for thinking humans} indeed. See \cite{chomskyChatGPTNYT2023,wallaceWelsAIslopNYT2024,mitchellAI2019,kobieLongHistory2024,julia2020there,sumpterFourWays2024}.

The purpose of this paper is twofold: to reveal the finesse of control system design beyond hoovering up indiscriminate uncurated data and passing it to the all-purpose `learning' algorithm; and, to study in some detail the combination of thought processes behind feedback control design in practice, including the links to control theory and to empirical science. Part of this analysis also is to highlight the role of simplicity needed in explanation and design, a theme of \cite{chomskyChatGPTNYT2023} and \cite{sumpterFourWays2024}, countervailing the rush to inexplicably greater complexity in models. The hope then is that we can differentiate between control applications amenable to benefit from machine learning and those inherently unsuited to it. Knowing the difference might actually be the intelligent way forward with increasingly powerful tools and methods, backstopped by an appreciation of the subtleties of dynamic modeling. Perhaps, a guide for thinking control engineers?

The methodology of control engineering is examined and interpreted within a context of the history and philosophy of science with some emphasis on the role of models, experiments and especially data. The objectives include appreciating the logical distinction between control theory and control practice, the role of deliberately designed experiments and data in feedback design, and the development of a philosophical basis for control design in practice. History provides a lens through which to appreciate the underlying philosophy of science and its development over the ages. The refinement of ideas concerning science \cite{Laudan:1996} and technology \cite{wootton:2015} is informative of how to view the engineering approach to feedback control design. Elements from the philosophy of science, notably deductive, experiment-design and inductive steps, are examined in principle and then with the benefit of applied industrial examples. This helps elucidate the roles of control theory, experiments and data in the feedback design process.

The paper attempts to segue repeatedly between the broad philosophical context and hard engineering examples. To instantiate ideas and add detail to vagaries, we incorporate a number of examples, each validated by their demonstrated commercial and industrial viability and each terminating with \textit{ein Blickwinkel} or perspective in German. As David Wallace-Wells \cite{wallaceWelsAIslopNYT2024} poses, before investing hundreds of billions of dollars we really ought to ask what is the trillion-dollar problem which might potentially be solved. By sticking to industrially proven technologies, we hope to delineate what works suitably well in practice or was judged worthy of the risk.

\section{Intentions of control}\label{section:Intention}
In Control~101, we learn that feedback can alter the behavior of the open-loop system in three principal substantive ways.
\begin{itemize}
\item Open-loop unstable systems can be stabilized or system dynamics otherwise altered.
\item The effects of external disturbances on the closed-loop output signals can be mitigated.
\item The closed-loop dynamics can be rendered less sensitive to variations in the open-loop dynamics.
\end{itemize}
In University of California Berkeley's Multivariable Feedback Control EE221B of 1972, this list is refined and augmented to include the following \cite{callierDesoer1982}.
\begin{itemize}
\item Asymptotic tracking can be achieved.
\item A nonlinear open-loop system can have improved linearity in closed-loop.
\item Instability can be achieved for oscillators and the like.
\end{itemize}
More could be added, such as decoupling for simpler design and comprehension.

Likewise, in Control~101 the direct elementary design and implementation schema is: build a linear time-invariant model; apply the linear control design tools; implement the feedback controller in electronics; tweak or optimize; walk away. However, beyond simple examples such as the DC servomotor example later, this schema usually falls at the first and many subsequent hurdles. Myriad design options: for the system configuration in terms of sensor and actuator selections and placements, altering the achievable behavior; operating regime affecting the appropriate model; model structure being central; realization problems arising; lack of clarity of which among the competing interests and objectives is most important. \AA str\"om's admonition is only one of many impediments to, or, euphemistically, `design choices' for, achieving control. Indeed, the design side of control is both its attraction and its workload. As increasingly non-ideal behavior is admitted the more difficulties can arise and the control objectives recede in their achievability. Specific examples follow later.

Early on, especially early 20th century, these ideas of control performance were quantified in statistical terms through graphical time-series tools such as the Shewhart Chart, a chart recording of controlled (i.e. output) process variables over time accompanied by visual tests to determine whether the process was ``under control.'' These tests can be related to deviations of mean value or standard deviation or change in the whiteness of residuals. This is the realm of Statistical Process Control, which has evolved into a range of tools for management in quality control and process improvement, such as Six Sigma. The importance for us is that it establishes the fundamental benefits sought of feedback control, its basis in modifying system dynamics and assessment via statistical or probabilistic variability as is seen in the measurements. Our aim will be to concentrate on the control engineering side, as opposed to management or process improvement, but to preserve these ideas of utility, purpose and assessment, primarily the role of variability in the environment or the plant itself. This is not to diminish the importance of the control engineers appreciating the overarching process objective of their own role.

The apparent rush to reliance exclusively on data to achieve closed-loop control is recent and reflects the currently prevailing technology milieu and publicity. Although, industry has also shown a sharp interest in these apparently low-overhead hypothetical solutions. Appeals are made to Willems Fundamental Lemma \cite{MarkovskyWillemsVanHuffelDeMoor:2006}, which relates to noise-free linear systems of known state dimension, even though it is unclear how these conditions might relate to control solutions addressing the problem classes above; themes abiding throughout this article. The proclaimed deluge of data tracks the growth of the internet and general accessibility to material by search \cite{doomenJInforEthics2009}. It does not necessarily pertain that industrial process data are somehow more prevalent or, especially, informative -- indeed, data archiving has long been an important and expensive aspect of engineering information technology. Although recent capabilities include being able to access data from disparate processes and sites. There has, however, been significant advance in study and capabilities from machine learning, albeit not apparently in safety critical sectors. It is sobering how, after so many decades of effort, so few adaptive control systems are in operation in engineering systems beyond telecommunications, where data informativity for agile adaptation is achieved at significant expense in throughput \cite{LeeBahkMidamble2021} and modest reliability.

The critical but less immediate feature of these raisons d'\^etre for control above is that the assessment of success takes place in an operating environment distinct from that in which the system is initially manifested, since the presence of a candidate feedback controller will intentionally alter the closed-loop dynamics and responses. The role of data, collected in one operating environment but intended for performing in a different environment, is central to our concerns. 

Before embarking on a philosophical journey, let us first agree on the terminology. See \nameref{sidebar-glossary}.

{\section*{Sidebar: Agreeing to terms}\label{sidebar-glossary}
Since the subject matter necessarily intersects the philosophy of science, it is instructive to define some terms, a number of which are based loosely on the New Oxford American Dictionary. Where terms, are imprecise or ambiguous in their usage such as `data-driven,' we shall restrict their meaning to that defined here. We do this for clarity.
\begin{description}
\item[Epistemology] --- The theory of knowledge, especially with regard to its methods, validity and scope. Epistemology is the investigation of what distinguishes justified contention from opinion. 
\item[Deductive reasoning] --- inference in which the conclusion about particulars follows necessarily from general premises or hypotheses.\\
\item[Inductive reasoning] --- inference of a generalized conclusion from particular instances. This includes, for example, fitting a model of given structure to data or the selection of model structure from data.
\item[Science] --- the systematic study of the structure and behavior of the physical and natural world through observation, experimentation, and the testing of theories against the evidence obtained.
\item[Empirical science] --- is a synonym for science, emphasizing experimental evidence. Empiricism is a branch of epistemology positing that knowledge comes only from tangible evidence.
\item[Experience] -- the collection of observational evidence informing or prejudicing the subsequent scientific experimental analysis. 
\item[Data-driven methods] --- operate solely from the available data without connection with the data generating process and its design. That is, these methods are purely inductive for the purposes of this paper.
\item[Engineering] --- The application of scientific and mathematical principles to practical ends such as the design, manufacture, and operation of efficient and economical structures, machines, processes, and systems.
\item[Instrumentalism] --- a doctrine that ideas are instruments of action and that their usefulness in application determines their validity\\
\item[Truth, belief, real] --- not words we use in a technical sense.
\end{description}
}

\section{The epistemology of control}\label{section:Epistemology}
A control engineer might ask several preliminary questions before embarking on a design
\begin{quote} \textit{What do we want to achieve?}\\\textit{Which available system design choices might facilitate this?}\\ \textit{What do we know about the plant, its environment and its variability that might allow us to do this?}
\end{quote} The epistemological question would be also to ask 
\begin{quote}\textit{How do we know these things about the system? And how do we distinguish knowledge from belief?}\end{quote}
And the scientific question might follow 
\begin{quote}
\textit{What experiment might we perform on the system to endeavor to falsify or corroborate this presumed knowledge?}
\end{quote} 
And the engineer returns to interpret the response to each of these latter questions or results of experiments solely through the lens of control intention and demonstrated achievement. This introduces a central theme regarding the relationship between knowledge, modeling and data, which is that each of these is accompanied by a confidence level concerning the utility and reliability of this information for the purpose of feedback control design and performance in application. In the end, performance assertions can only be validated via implementation and measurement.

It needs to be emphasized that control design embodies much more than control law design, that is, the selection of feedback as a dynamic function of measured signals. It includes system selections for actuators and sensors, their specification and placement. Above this system level, it also encompasses determination of the economic and operational requirements and objectives. However, since our focus here is primarily on the aspects dealing with data prerequisites, attention will largely be on these issues, but within the commercial context.

\section{The philosophies of science and control}\label{section:Philosophy_of_Science}
Data, experience, prior knowledge and art all play a role in the formulation of feedback control design; they help us to decide which theories or associated design methodologies might yield candidate controllers. The acceptance and commissioning of a particular feedback controller relies on closed-loop experiments and analysis, usually statistical, of this data, since \textit{ab initio} guarantees of performance are products of hypothesis and deduction alone. In comparison to control theory, this latter testing stage is reliant on inductive reasoning and the probabilistic determination of functionality in closed-loop. This interplay between three stages -- hypotheses, experiment, evaluation -- undergirds the controller design and implementation process. It draws on many disciplines particularly control theory, experiment design and statistical hypothesis testing. This section is focused on the placement of these ideas in a milieu of science, where philosophers, historians and scientists themselves have had much longer to assess why they do these things.

The philosophy of science is a longstanding, evolving and robustly debated active subject. See Curd, Cover and Pincock \textit{Philosophy of Science: the central issues} \cite{CurdCoverPincock2012} for a comprehensive historical review containing many of the core source papers from the major contributors to the field. The subject is often inseparable from the history of science \cite{wootton:2015}, indicating the changing perspectives over time and the involvement of contemporary scientists in declaring how and why they themselves pursue their subject. For our interest in the philosophy of feedback control, the clear connections to scientific methods, the well developed ideas and copious but well-organized literature of science provide a useful environment from which to consider our specific questions. Science also is a foil to control in considering how control might differ, if at all, from purely scientific pursuits. Recognizing the multitude of varying positions on science, there could be many acceptable descriptions of the philosophy of control. This paper emphasizes the author's own position as a person more experienced and skilled in the control side of the business than the philosophical. The science philosopher closest in viewpoint to theirs is Larry Laudan \cite{Laudan:1996} whose emphasis is on the utility of science, eschewing concepts such as reality and truth which can be malleable over time. The usual utility of science is its capacity to predict outcomes in new and challenging experiments. Utility for a feedback controller is outlined in the section above on the intention of control and centers on the achieved performance in application. This is a point of distinction between control and science.

A general view of scientific method, and indeed of empirical science in general, is that it proceeds from a combination of steps involving hypotheses, deductions, experiment design and data gathering followed by inductive refutation or corroboration of the hypotheses. That is, many elements of reasoning are involved and coordinated in sequence, as revisited iteratively in the light of new information. Inductive science breaks from this in deemphasizing the deductive aspects. Purely data-driven methods would apparently eschew the deductive step completely. Such approaches, with their catchcries of  `Let the data speak for themselves!' and without regard to the provenance and nature of the experiment yielding the data, might be classified as inductivist, since they appeal solely to induction as a path forward to extract knowledge, predictions or conclusions. Aspects of machine learning and data-driven control fall into this classification. And it is the aim here to tease out conditions when this might or might not be warranted. This is briefly discussed in \nameref{sidebar-induction} and distilled in the conclusion.

Evidently, in control systems the slew of papers reliant on induction-only approaches must be subservient to the notion for induction about uniformity of systems before and after the current time. This might be compared to the reliance on models in control, where the model -- structured by deduction and fitted by induction -- is presumed to capture variations in operating conditions. This is akin to an interpolation-extrapolation issue. As is addressed later, the problem in control rests on the observation that alteration of the controller can (hopefully) lead to non-uniformity of operating conditions. Models can be construed as an effort to capture system response to changes in the operating environment.

Karl Popper's \cite{Popper:34,popperRefut1962} focus on the scientific method and \textit{Falsificationism} is one in which phenomena in the physical world are the subject of predictive hypotheses and novel demanding experiments are designed to try to refute these hypotheses. Unfalsified (or \textit{corroborated}) hypotheses persist for future assail and adoption into the current version of science. The emphasis in these experiments is that they should be significant stress tests for the proposed ideas and not simply validation tests based on repeating the same experiment. These ideas of unfalsified models have also been advanced in a control context \cite{Kosut:95,Safonov&Tsao:97}. Popper suggests that the selection of hypotheses can be unencumbered from justification provided that a suitably rigorous experiment is proposed for their refutation.  Feynman provides an affirming view.
\begin{quote}\textit{Now I'm going to discuss how we would look for a new law. In general, we look for a new law by the following process. First, we guess it (audience laughter), no, don't laugh, that's the truth. Then we compute the consequences of the guess, to see what, if this is right, if this law we guess is right, to see what it would imply and then we compare the computation results to nature or we say compare to experiment or experience, compare it directly with observations to see if it works.\\
\\If it disagrees with experiment, it's wrong. In that simple statement is the key to science. It doesn't make any difference how beautiful your guess is, it doesn't matter how smart you are, who made the guess, or what his name is ... If it disagrees with experiment, it's wrong. That's all there is to it.}
\flushright{Richard P. Feynman \cite{FeynmanMessenger1964}}
\end{quote}
In practice of course, we are guided by experience from early experiments, process insights from design, availability of theory, and other factors, along with guesswork and conjecture. That is, our hypotheses are informed by prior experiments and empirical data. This involves hypothesis and deductive reasoning capable of rationalizing observations. There is quite some span in the discussion of the permissible sources of conjectures and hypotheses. Appeal to deities is, however, not advised.

In science there is significant appeal to simple conjectures capable of explaining the data. Occam's razor enunciates this and the approach has been formalized in Solomonoff's theory of inductive inference \cite{solomonoff1964-1,solomonoff1964-2} based on the extrapolation of data sets and building on Kolmogorov's complexity \cite{kolmogorov1983}. The source quote, often misstated, from Einstein on the matter
\begin{quote}
\em It can scarcely be denied that the supreme goal of all theory is to make the irreducible basic elements as simple and as few as possible without having to surrender the adequate representation of a single datum of experience \cite{einsteinQuotes:2013}.
\end{quote}
captures the challenge of hypothesizing from data while preserving parsimony. In system identification, Rissanen \cite{rissanenAutom1978} develops related complexity measures for determining model order. The upshot of this is that it is difficult to formulate rules for the inductive data-oriented component, which in part reflects the variability of empirical data and builds in experience. The scientific process comprises alternating stages of: conjecturing predictive explanations; deducing consequences stemming from these conjectures; and formulating demanding empirical tests to falsify or corroborate the conjectures. Note that \textit{a single datum} above is likely hyperbolic or statistical given the nature of evaluation.

One systematic but detailed view of science, discussed later, is that science involves proposing hypotheses concerning the unobservables of phenomena, deducing their consequences for the observables of the system and then testing for this. Think, for example, of unobservable spin of individual electrons in a material and the observable bulk property of magnetic field. Much of the philosophical study of unobservable phenomena implying observable behavior lies in the problem that the observable data do not determine a unique set of rules for the unobservable phenomena; they are underdetermined and therefore allow for many potential explanations. The language and concept of observability is familiar to control. Although the effect of experimental conditions in changing observability is less well appreciated \cite{LiuBitmead_Autom2011}.

\subsection{The Philosophy of Technology versus that of Science}
Philosophers have long sought to appreciate the similarities and distinctions between science and engineering or technology. The utility-driven focus of engineering and its objective to alter the world, rather than simply to describe it as it is, seem to provide a difficult separation. The emphasis on design is, however, a common theme in efforts to appreciate the distinctions. Whence,
\begin{quote}\textit{Technology or engineering as a practice is concerned with the creation of artifacts and, of increasing importance, artifact-based services. The design process, the structured process leading toward that goal, forms the core of the practice of engineering.}\cite{sep-technology}
\end{quote}

Vincenti \cite{vincentiBook1990} delves into these contrasts using a sequence of examples from application-oriented aircraft design and problem solving approaches. He goes further to categorize the many aspects of engineering knowledge and epistemology, striving hard to separate engineering from simply being applied science. The motivation and tools are distinct in his formulation. That being said, there is a strong commonality between science and technology, with each dependent on the other at the applications end.

Control shares both the technology and artifact-based services background together with the significant overlap with scientific methods. Indeed, control would seem to be in mutual heavenly motion about science, technology and mathematics, and thereby the common elements of their philosophies. The design emphasis is moderated by specifications and hence can yield a terminating effort and concentration on satisficing versus optimality \cite{satisficingTSMC1998}.

\subsection{Control versus science}\label{section:Control_vs_Science}
How does feedback controller design essentially differ from science in general? 
\begin{itemize}
\item The objectives of control laid out above are focused on achieving closed-loop behaviors which are not necessarily linked to predictive performance in every novel environment. Although, for some system classes and objectives, prediction and control performance are related \cite{harris1989}.
\item In control applications, there is always evaluation of performance using measurements from the target system. In physics and notably cosmology, it is not always feasible to conduct empirical evaluation of methods and ideas \cite{evaHartmannSynthese2021}.
\item In control design, we would appear to be more than usually constrained by the limited set of design tools and their capacity for accommodating complexity. 
\item Middleton and Goodwin \cite{MiddletonGoodwinBook1990} suggest that control systems are restricted to two frequency decades of bandwidth in their conception and for their operation. This is the framework for their numerical $\delta$-transformation and is a differentiator between control and, say, wideband signal processing.
\item Allied with the two preceding points, there is an impetus to keep modeling very simple in order to be able to apply design tools affecting only the dominant dynamics. Here we definitely can be at variance with Einstein, when, say, high-frequency data fit is immaterial for control performance.
\item The availability of commercial systems for experimentation can be highly variable and expensive. Low-value products might be wasted or recycled while high-value processes might present cost or availability problems. So, economic considerations can impinge on the experimentation phase, which is a serious limitation in control practice. Of course, cost constraints affect science experiments too but are decoupled from commercial operating costs.
\item Control design is a terminating procedure within a time and human budget. Alf Isaksson of ABB Corporate Research suggests that the final hotel bill of the commissioning engineers of a control system is a core figure of merit for performance.
\end{itemize}

How might control design conform with the scientific method?
\begin{itemize}
\item Model-based control design follows from the empirical fitting of a dynamic model using an experiment followed by system identification. System identification itself entails the informed selection -- some would say conjecture -- of model structure and the design of a suitably informative experiment for its fitting \cite{Goodwin&Payne:77}.
\item Many controller tuning rules, such as open-loop Ziegler-Nichols PID, rely on fitting deliberately approximate parametrized models, such as first-order-plus-time-delay. Equally, when applied in closed-loop they rely on intentionally causing, observing and calibrating oscillatory or other behavior. Such successive closed-loop experiments are closer to the refinements seen in the scientific method.
\item Iterative control design methods \cite{Zang&Bitmead&Gevers:95,Hjalmarsson&Gevers&Gunnarsson&Lequin:98,VRFTCampiLecchiniSavaraesi2002,AriyurKrsticBook:2003,VRFTexamples2023} consider a sequence of control designs, either model-based or ``direct.'' These approaches to successive controller revision use informative closed-loop experiments to adjust the controller for improved performance. Again, this is the interplay of experiment design and inductive modification of the controller, often based on gradient estimation methods deduced from process assumptions.
\end{itemize}
 
\section{The logic of control}\label{section:Logic_of_Control}
Control theory is essentially a deductive pursuit. (See \nameref{sidebar-CTiD}.) That is, assumptions and premises are defined and then their logical mathematical consequences are proven or deduced to follow from these hypotheses. The art of control theory is to develop formal problem statements with assumptions which are deemed likely useful for some class of significant practical domain and which are tractable for the proof. That is, the problem specifications should be both significant and soluble in order to carry weight. Establishing significance is frequently the sticking point. Although, correctness clearly is a central but subsidiary requirement. In this latter regard, it is important that \textit{all} assumptions are included or traceable \cite{PolyaHowToSolveIt1957}. It is interesting to note that, according to this narrow view of control theory, it has no contact with the inductive use of experiments and data.

By contrast, control practice involves data, experience, prior knowledge, experimentation, induction and deduction, and quite some art. In this regard, it has much in common with empirical science, to distinguish it from mathematics. Accordingly, aspects from the scientific method and its attendant philosophies can guide thinking. Where control deviates from science lies in the overarching objective of utility for control application and not simply general predictive power of explanations. And our control design tools can be quite limited in scope or, equivalently, exacting in their restrictions on the properties of the plant, such as linearity, passivity, open-loop stability, etc. Thus control design involves a very constrained blend of simplicity and utility. The advantage, however, is that feedback controller design can avail itself of closed-loop data, which in iterative design schemes \cite{Zang&Bitmead&Gevers:95,Hjalmarsson&Gevers&Gunnarsson&Lequin:98,VRFTCampiLecchiniSavaraesi2002,AriyurKrsticBook:2003,VRFTexamples2023} approximates the eventual operating environment of the controller. The \textit{quid pro quo} of this aspect of control is that, as the closed-loop changes, there is no requirement for new models to subsume or maintain consistency with their antecedents. Likewise, past data can become deprecated.

{
\section*{Sidebar: Control Theory is Deductive}\label{sidebar-CTiD}
Control theory is anchored in mathematics and the formal derivation of system properties as logical consequences following directly from the assumptions. Like the rest of mathematics, control theory is not an empirical science since it has no reliance on experiment or data. So, control theory does not use inductive reasoning. The role of empiricism and induction for control theory lies in the assessment of the realism and suitability of the assumptions being advanced and in the consideration of the potential subsequent utility for feedback control of the predicate properties. 

This is not to say that control theory is silent on data and experiments. It is the source and support for many tools. System identification is an important aspect of modeling for control linked in theory to statistics. The theory leads to methods for analysis of empirical data but without limitation to specific data. The assumptions, such as stationarity or asymptotic independence, are the basis for formal proofs of consequences, i.e. deductions. The application in practice rests on the (approximate and local) validity of these assumptions or premises.

Guarantees and certificates of performance belong to the realm of theory, since only deductive methods are capable of delivering such absolute surety. In practice, it is in general not possible to verify the satisfaction of the premises without detailed incontrovertible system knowledge, which must extend to knowledge concerning the disturbance signals active during data collection and later operation. Data-driven methods with guarantees blend deductive assurances with induction based on data. They hence depend on warranties on the experimental data, which renders them not fully inductive. This distinction between deductive mathematics and inductive science is a topic of central debate in the philosophy of science.
}

{
\section*{Sidebar: Induction and (purely) data-driven methods}\label{sidebar-induction}
Inductive reasoning starts with the data and moves to general conclusions. Purely inductive methods do not engage with the provenance and generation of the data but limit themselves to drawing inferences from the data alone. Such a view of induction has invited critics at least a far back as David Hume in 1739 \cite{sep-induction-problem}, with the principal objection being that induction presumes the ``Uniformity Principle.'' That is, that future data will look, at least statistically, similar to past data. Without such a principle, the predictive capacity of induction cannot be advanced. Even with the principle, induction frequently requires a probabilistic framework. There have been many proponents \cite{stoveInduction1986} and opponents \cite{gianquintoStoveReview1987} of purely inductive methods, with some focus applied to the background assumptions undergirding the approach.

Despite its obvious logical appeal, induction has a chequered place in the philosophy of science \cite{sep-induction-problem,saatsiPhilOsci2005}. When one considers more recent inductive methods, such as machine learning and artificial intelligence, Hume's problem abides and is met with the proclamation that the data are presupposed to be \textit{representative} without other evidence, since the data are all that exist. Yet, this is difficult to verify or validate, particularly in the context of feedback control design to alter closed-loop system dynamics  deliberately. In system identification, which operates under quasi-stationarity assumptions \cite{Ljung:99b}, experiment design is in principle applied to ensure that the generated data suffice for reliably fitting parametric models of a given order; this is related to the Cram\'er-Rao Lower Bound on the variance. In turn, one may appeal to examination of the data to evaluate the suitability of the data set for such fitting \cite{coulsonEtAlQuantiativeWillems2023,berberichEtAlPEACC2023}. However these intrusions into the data generating and evaluation mechanisms and into models depart from induction and its hands-off approach to the data. 

From the perspective of this paper, reliance on inductive methods alone is advisable only in the case with prior knowledge that the data are and always will remain representative. It is also the thesis that this often is not the case, and especially so when designing and implementing feedback controllers.
}

\section{Models and model-based control}\label{section:Models}
\begin{quote}
\textit{``Since all models are wrong the scientist cannot obtain a `correct' one by excessive elaboration. On the contrary following William of Occam he should seek an economical description of natural phenomena"} George E.P. Box \cite{boxAllModelsJAmStatSoc1976}.
\end{quote}
Models have a long, involved but treasured place in the philosophy of science \cite{sep-models-science,FriggHartmannEncyclopedia2006}. They also provide an entr\'ee to model-based control design and therefore require either of: little introduction, or a deep dive into full philosophical reassessment. Three hardware model-based control designs will be considered in four of the seven Sidebars. There are many facets of models in science, as explicated in the above references. In control, models are largely idealized, stylized or simplified representations of the target system, in which the model is recognized to be a deliberate distortion. The advantage of control with models is two-fold: that feedback can be capable of diminishing closed-loop sensitivity to quantified model inaccuracy and that a robustness theory exists \cite{LiuYaoRobust2016}. Although, the ouroboros-like notion of \textit{uncertainty} enters which, in turn and like everything else, requires some level of certification. Once a model is obtained, and provided it is presented in an appropriate form, standard computational tools can be applied to design the model-based feedback controller.

Specifically with respect to controller design in practice, we have the opening quote from \AA str\"om, who also gave us the epithet \textit{Control, the hidden technology,} the title of the lecture from which this quote is sourced. \AA str\"om's remark, in particular, is germane in emphasizing that the development of a plant system model is probably the hardest part of model-based control applications, including implicit models as might drive PID control design. These are epistemic questions at heart, since the model embodies contentions about the target system and the model's capacity to deliver an acceptable controller via the design process. Arriving at a suitable model involves science and deductive logic together with experience, experiment and inductive logic. The art, craft or guile of control is to navigate the hypothesis development for these iterative steps. 

One might expand upon \AA str\"om's  statement by including that the model needs also to be fit for purpose for control design and therefore adapted to the eventual closed-loop operating environment. This will be amplified in the sugar mill control example to come. Where control design is explicitly model based, the limitations of the control design process are additionally affected by the model quality in replicating the behavior of the plant system both in open and in closed loop. This is further exacerbated by needing to dovetail the model inadequacies with the control design foibles \cite{AlbertosSala:02}. 

An intriguing feature of control design is the necessary partitioning of the measured signals into causally related input signals, output signals and disturbances, since this is part of the devising of control design. Behavioral models \cite{PoldermanWillems1998}, stemming from circuit theoretic ideas of interconnection without buffering at ports or interfaces, and other models flowing from coupled systems in thermal, mechanical or biological arenas do not immediately lend themselves to control design and indeed often proscribe the above partitioning. However once inputs and outputs are delineated, behavioral methods have been coopted into system identification in the noise-free exact modeling framework because they come with an attendant data adequacy condition for identifiability; the so-called Willems' \textit{Fundamental Lemma} \cite{WillemsRapisardaMarkovskyDeMoorSCL:2005}, first denominated as such  in \cite{MarkovskyWillemsVanHuffelDeMoor:2006}. The need to identify inputs, outputs, disturbances and causality is an integral part of control and also of the model building inductive parts of system identification, time series analysis or machine learning. How we know which signals are of each type and relation is an epistemological question, which is particularly apparent in econometrics \cite{Shojaie:2022aa}.

\begin{figure}[h]
\begin{centering}
\includegraphics[width=0.45\textwidth]{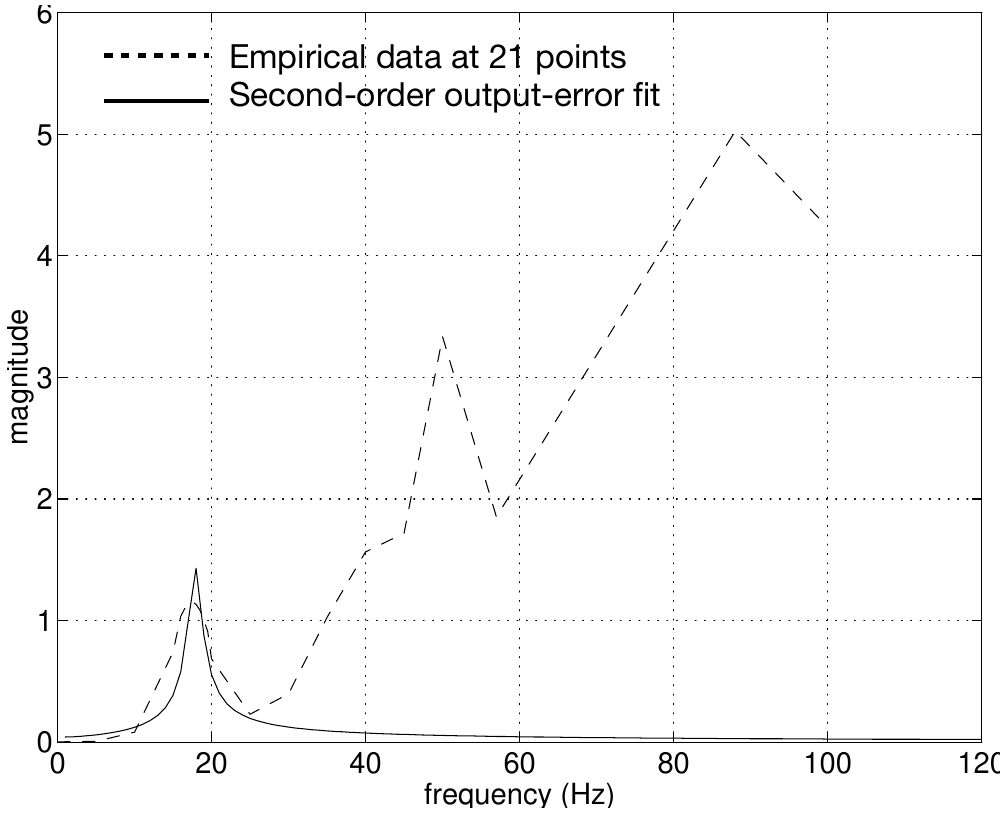}
\includegraphics[width=0.48\textwidth]{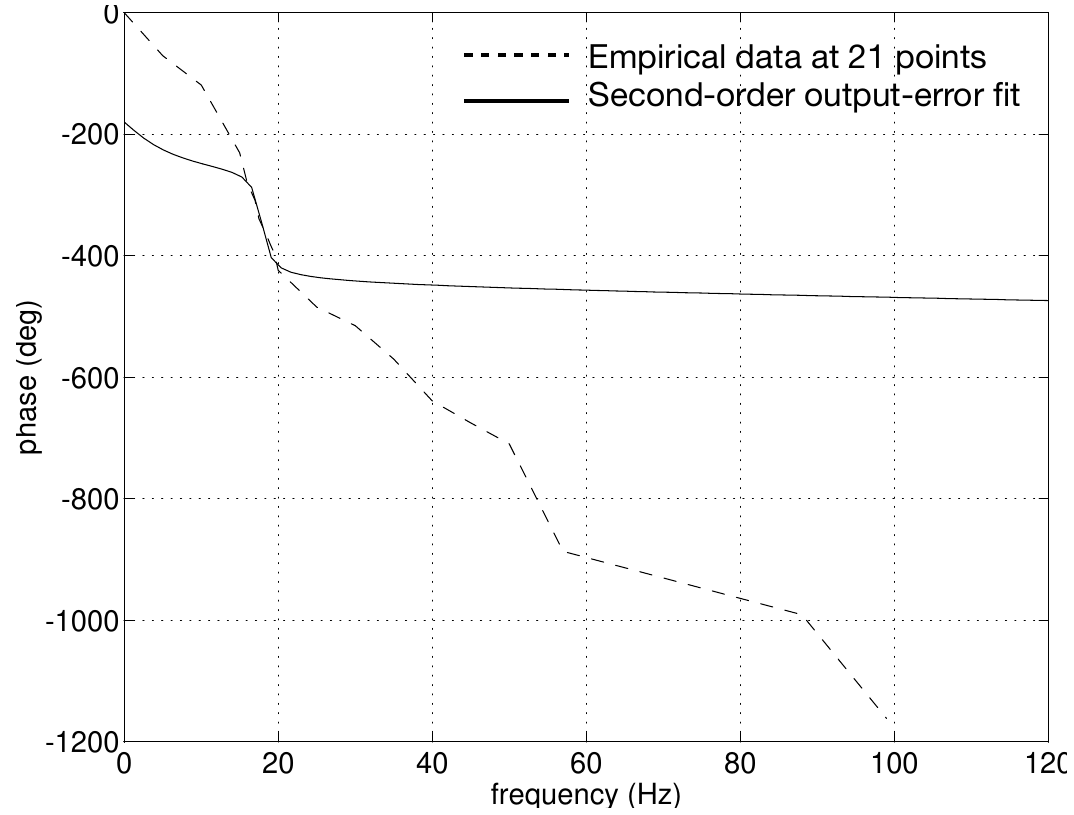}
\caption{\label{fig:heli}Helicopter vibration cancellation system modeling: 21-point empirical frequency response (dashed line) and fitted open-loop-stable second-order simplified model, from \cite{Crisafulli&BRJC:94}. The tight concordance in phase between experiment and model in the disturbance band of 16-18 Hz, together with open-loop stability of the plant and model, implies that this model is well suited to vibration rejection control design.}
\end{centering}
\end{figure}

Figure~\ref{fig:heli} shows, in the frequency domain, a model fitted to a helicopter vibration suppression system; voltage input to the active hydraulic damper valve to output measurement of the vibration accelerometer. The dashed curve is the empirically measured frequency response, using first a slow sine-sweep frequency analyzer and then excitation at 21 distinct unevenly-spaced frequencies followed by removing harmonics and computation of complex gains. The solid curve shows a low-order frequency domain fit, guaranteed stable by using output-error model structure. The vibration disturbance due to the helicopter rotor is a sinusoid of slowly varying frequency between 16.15-17.85 Hz. Active control is applied to diminish the vibration. From the perspective of the control problem, the model is close to ideal, since it captures the open-loop system stability and the phase requirements for disturbance rejection well \cite{Crisafulli&BRJC:94}. Additionally, it admits a closed-loop robustness analysis of the controller \cite{helicopterChapter2002}. The purpose of this example is to amplify the peculiarity of feedback controller needs of a model. We certainly are not seeking a physical model capable of predicting the output in any circumstance other than possibly with a sinusoidal excitation at a nominal frequency within the given range.

System identification is a mainstay of modeling for control design \cite{Ljung:99b}, particularly with finite-dimensional linear models. It describes both the inductive approach to the fitting of models using optimization based on data from experiments and the formal framework of quasi-stationarity of signals, model structures, random noise and second-order statistics. It also encompasses the adequacy of data for the fitting purpose via excitation and the design of experiments; these latter conditions being tied to identifiability or the uniqueness of the parameter which minimizes the (least-squares prediction-error) criterion. The (deductive) mathematical setting of statistics provides the environment for formulating the hypotheses for the inductive analysis. Outside of the strictures of system identification, there remain, perhaps unstated, assumptions about the target system and the signal properties. These might include linearity, known finite dimension, absence of noise, adequacy of the data for identifiability, measurement quality, etc. The central point here is to recognize that modeling from data involves the same scientific tools of hypotheses, deductions, experiment design and induction which comprise much of science. It is the way these elements are combined and for what purpose that specializes in modeling for control.

\section{Observables, unobservables and states}\label{section:Observables}

Scientific models frequently appeal to subdivision into observable or unobservable phenomena. For the statistical theory of gases, think pressures, temperatures and volumes for a gas as observables and the motion of individual molecules as the unobservables. Or, for chemical process control, the measurements from the reaction versus the detailed internal dynamics within the reaction vessel. For control systems people, these are quite familiar concepts and, indeed, physicists often speak of the state of a system as capturing such unobservables, even though control people have a much more reserved usage of the word \textit{state} \cite{rosenbrockState99}. Raginski \cite{raginskyState2024} provides a fuller analysis of state in control parlance and of its historical connections. Models in science connote the development of descriptions of the unobservables which manifest in behavior of the observables. This is a useful viewpoint for control as it encapsulates the elements (some might say imponderables) of model inaccuracy, external disturbance evolution, plant variability, etc into these unobservable components. It does, however, forestall the precise measurement of the state of a plant in practice; notably for adherents to Box's comment above, where the state of the model is something of a fiction. This brings into question for these target systems the formal notion of state at time $t$ as a sufficient set of information at time $t$ so that, given the current state and all future inputs for $\tau\geq 0$, we can exactly calculate all future outputs at times $t+\tau$. The disjunction between model and target carries over to that between model state and target system state, were it actually to exist.

This raises the question of whether, in the inductive world of dealing with data, the target system state can be measured precisely. The state is a speculative construct from the deductive world. Methods involving state derivative feedback \cite{LewisSyrmosTAC1991} move further from realizability, since it is uncertain how such information might be constructed from data. While seeking to generalize output derivative feedback from PID control, these methods fail to recognize the practical requirement of low-pass filtering within any derivative action \cite{AstromMurrayBook:2016}. In application \cite{DuanNiKoCACIE2005}, this filtering is part of the state derivative estimator. Nevertheless, in control theory it can be helpful to speak of the state (or even state derivative) as being measurable. The issue comes, front and center, in control practice.


%
{\section*{Sidebar: DC servomotor identification and control}\label{sidebar-mip}
A DC servomotor is the actuator for a self-balancing demonstrator robot MiP (\textit{Mobile Inverted Pendulum}) marketed by WowWee, shown in Figure~\ref{fig:mip}. To build a feedback controller for the motor, a finite-dimensional linear model is fitted to experimental data. The experiments were conducted with a MiP in repose, that is lying on its back, since the erect system is unstable.

\begin{figure}[h]
\begin{centering}
\includegraphics[width=0.5\textwidth]{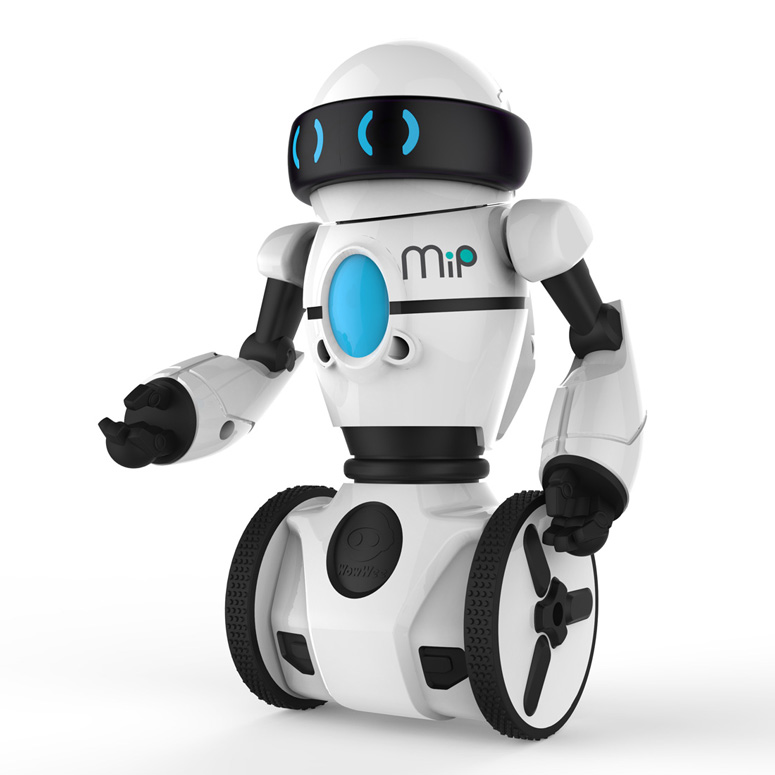}
\caption{\label{fig:mip}WowWee MiP robot.}
\end{centering}
\end{figure}

Sixty frequency response experiments were conducted, voltage input to shaft encoder angle, with sinusoidal frequencies from 0 Hz to 45 Hz and sampling interval of 5 ms. From each data set, the frequency response was computed after allowing for transients to fade. The empirical Bode plot was constructed and used to fit models of successive complexities. This is displayed in Figure~\ref{fig:mipBode}.

\begin{figure}[h]
\begin{centering}
\includegraphics[width=0.75\textwidth]{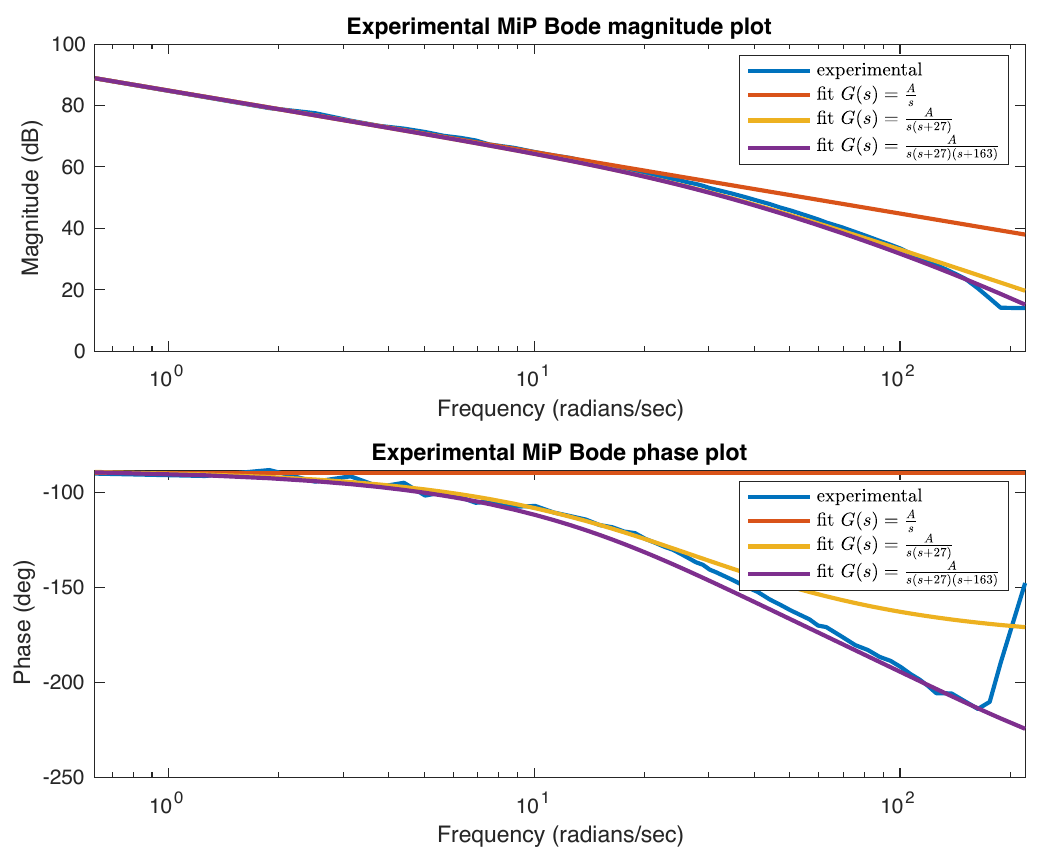}
\caption{\label{fig:mipBode}Empirical Bode plot of MiP with successive model fits.}
\end{centering}
\end{figure}

The second-order fitted model, $P(s)=\frac{A}{s(s+27)}$, was used to design a successful feedback controller to stabilize MiP and to develop a path tracking approach. In closed-loop operation, this controller yielded satisfactory performance.

\subsection*{Ein Blickwinkel} 
The servomotor control design and implementation proceeds as follows.
\begin{enumerate}[label=\roman*.]
\item The electromagnetic physics of DC servomotors, based on Maxwell's Equations, themselves idealized from empirical observations, yield linear dynamics linking input voltage and rotor angle. Manufacturer datasheets derived from experiment confirm and quantify the close-to-linearity of this target system under the operating conditions. This is the epistemological starting point.
\item Under the hypothesis of second-order linear dynamics of given structure parameterized by two parameters and operating the motor with bandwidth below 50 radians/second, any experiment comprising at least two complex sinusoids of distinct frequencies should suffice for fitting the linear model. This is a deductive step.
\item The experiments were conducted and the dynamic model of the given structure fitted using this data. A validation test was performed using excitation up to 50 radians per second, empirically corroborating the model fitting hypothesis. This is the inductive step.
\item A linear state feedback controller was designed with the bandwidth in mind, deductive, and operated well; induction again. 
\item There are no safety critical aspects of the control.
\end{enumerate}

In a sense, this example is the archetype of model-based controller synthesis: conduct an experiment, collect data, fit a model, design the controller, implement and test. The first three steps could be omitted by using the datasheet and detailed calculations of moment of inertia etc. Although, it is important to note the requirements imposed on the experiment by the model structure; that is, that any sufficiently rich experiment would suffice.
}

{
\section*{Sidebar: FrothSense+ an inductively trained instrument}
Flotation separation of mineral-bearing parts of finely ground ore slurry is one of the most economically important processes in metals mining \cite{willsMinerals2016}. It relies on differing hydrophilicity between valuable components and gangue minerals in a water-reagent mixture, with different reagents specified to target particular minerals. A froth of bubbles is generated by injecting air into the water/slurry mix, or \textit{pulp}. The bubbles float to the top of the cell and the valuable minerals, conditioned by the reagent to be hydrophobic, attach to the bubbles as they rise. So the product is concentrated by preferential adherence to the bubble surface. The mineral-coated bubbles then either collapse, which rate is managed by other reagents affecting surface tension, or float to the surface as a froth which is then collected. A sequence of successive floatation cells each achieves a small improvement in concentration, say 3-5 times, and in mass yield, around 5-15\%. The sequence of cells achieves increasing product concentration and recovery with final separation being about 3-4\%\ of total feed in copper mining. 

Control of the flotation cells is based primarily on managing froth velocity and depth, as well as air flow rates, although the controllable resolution and cadence of the later depends on whether air is forced into the cell, or whether the cell is self-aspirated. A stable froth phase is the key to achieving high metallurgical performance. Other observables in the flotation process are the froth color, texture, and bubble size distributions. Feedback control of the flotation process requires estimation of froth velocity, color, texture, and bubble size distributions. Froth depth is easily measured with level instruments. Today, with relatively inexpensive and small networked cameras, that can provide high frame rates, estimation of these observables, using image analysis software, can readily occur in real-time and on a per cell basis. Within a large mineral processing concentrator, with modern sized flotation cells, there can be, say, 100 cells, sometimes more. Minimizing undesirable metal recovery losses with feedback control is fundamental to maximizing a business objective in mining.

Mining company Metso Minerals markets an industrial sensor, FrothSense+ (formerly VisioFroth), which uses illumination of the froth and camera images to provide online estimates of the key froth variables. This is described in \cite{yianatosMineralsEng2016,visioFrothSlides2007} and illustrated in Figure~\ref{fig:visioFroth}.

\begin{figure}[h]
\begin{centering}
\includegraphics[width=0.6\textwidth]{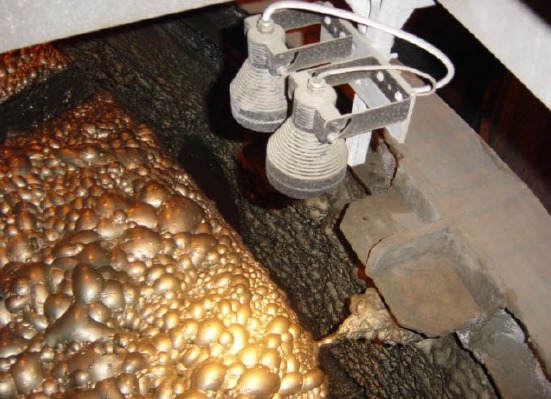}
\caption{\label{fig:visioFroth}Image of the illuminated froth within a floatation cell using VisioFroth, from \cite{visioFrothSlides2007}.}
\end{centering}
\end{figure}

For this mineral process, the sensing for recovery control is the difficult part of feedback control and FrothSense+ has made significant commercial impact and inroads. The image analysis is based on machine learning from copious data backed by laboratory confirmation. Statistical correlation analysis provides the validation of performance \cite{visioFrothSlides2007}. Explicit deductively derived models were eschewed in favor of data-driven induction. And the instrument works reliably and quantifiably well. How is this problem suited to purely inductive methods given their well-known philosophical problems and Hume's Uniformity Principle pr\'ecised in \nameref{sidebar-induction}?

\subsection*{Ein Blickwinkel}
Floatation separation is about one hundred years old, meaning that it predated digital automation but not visual human observation. It possesses these key features.
\begin{itemize}
\item Its operating environment is known to be statistically stationary, so that past data and future data are comparable and any sufficiently large data set should be \textit{representative.} 
\item Variations for grinding mill settings and mineral or reagent types can be accommodated via control system set point changes. The sensor will still provide actionable estimates.
\item Image analysis is a longstanding area of computing effort and application of machine learning to image classification has been well explored.
\item Image-based estimation of froth parameters would present a major task for a human coder to write an effective program.
\item The significant number of parallel and serial floatation cells and the statistical nature of the recovery and concentration process mean that the commercial performance is highly tolerant of estimation errors at the sensing stage and likewise responsive to statistical improvements in estimation.
\item Control law selection is outside the ambit of machine-learning aspects.
\end{itemize}
}

{
\section*{Sidebar: Combustion instability modeling and control}\label{sidebar-combustion}

\subsection*{The problem} Combustion instabilities are thermoacoustic oscillatory phenomena associated with low equivalence (fuel-to-air) ratio combustion where there is a nonlinear interaction between heat release and acoustics. They have been demonstrated experimentally using the Rijke Tube from the mid-1800s and have been the subject of active control. (See the excellent survey in \cite{AG-CSM-95}.) In jet engine combustors, these instabilities appear at low equivalence ratio, where it would be desirable to operate due to the corresponding low operating temperature with attendant benefits for nitrous oxides, NOx, emissions and for maintenance. However, these benefits are lost due to the oscillations between very-high- and very-low-temperature combustion. It is desired to build a control system to attenuate the oscillations in operation. 

\subsection*{Step 1: Initial experiments and data}
\begin{figure}[h]
\begin{centering}
\includegraphics[width=0.7\textwidth]{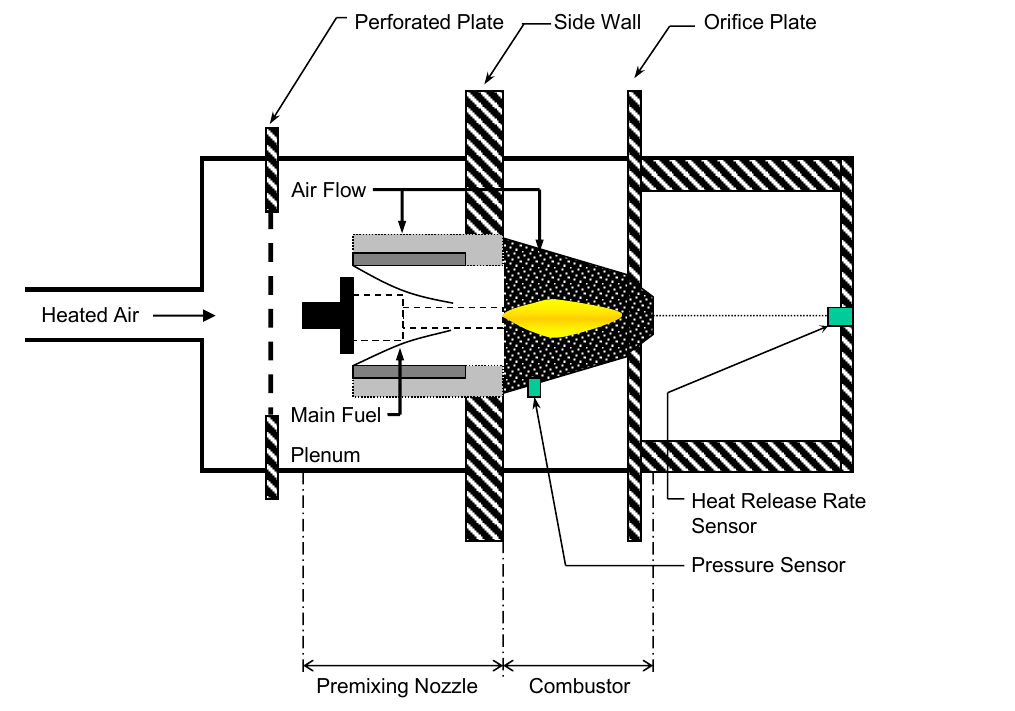}
\caption{\label{fig:snozzle}Elevation sketch of the UTRC/DARPA Single-Nozzle Rig mid-1990s in East Hartford Connecticut USA.}
\end{centering}
\end{figure}

In the mid-1990s an experimental apparatus was constructed at United Technologies Research Center, part of Raytheon since 2020, funded under a DARPA program. This is depicted in Figure~\ref{fig:snozzle}. The combustor is a single element from a ring of combustors such as might be found in an aircraft jet engine; United Technologies included Pratt \&\ Whitney Aircraft Engines. There are two sensors: a pressure sensor set at a slight remove from the combustion chamber to preserve it against heat; and, an optical pyrometric heat-release-rate sensor behind a protective window. A sequence of six experiments was performed at equivalence ratios, $\phi$, of 0.56 to 0.45 times the stoichiometric ratio of $\phi=1.0$, at which value all of the fuel and oxygen are consumed. Sustained oscillations were observed in each case with the coherence of the oscillations diminishing with $\phi$. Data were sampled for 3.2 seconds at 5 kHz in both channels during the ``stationary'' phase of the oscillations. 

Example data from the $\phi=0.56$ experiment are shown in Figure~\ref{fig:dataPlot1} with nominal units on the abscissae. The heat-release-rate is only positively valued and so its logarithm is plotted, since it exhibits a roughly symmetric sinusoidal form inviting close comparison with the pressure signal. 

\begin{figure}[h]
\begin{centering}
\includegraphics[width=0.5\textwidth]{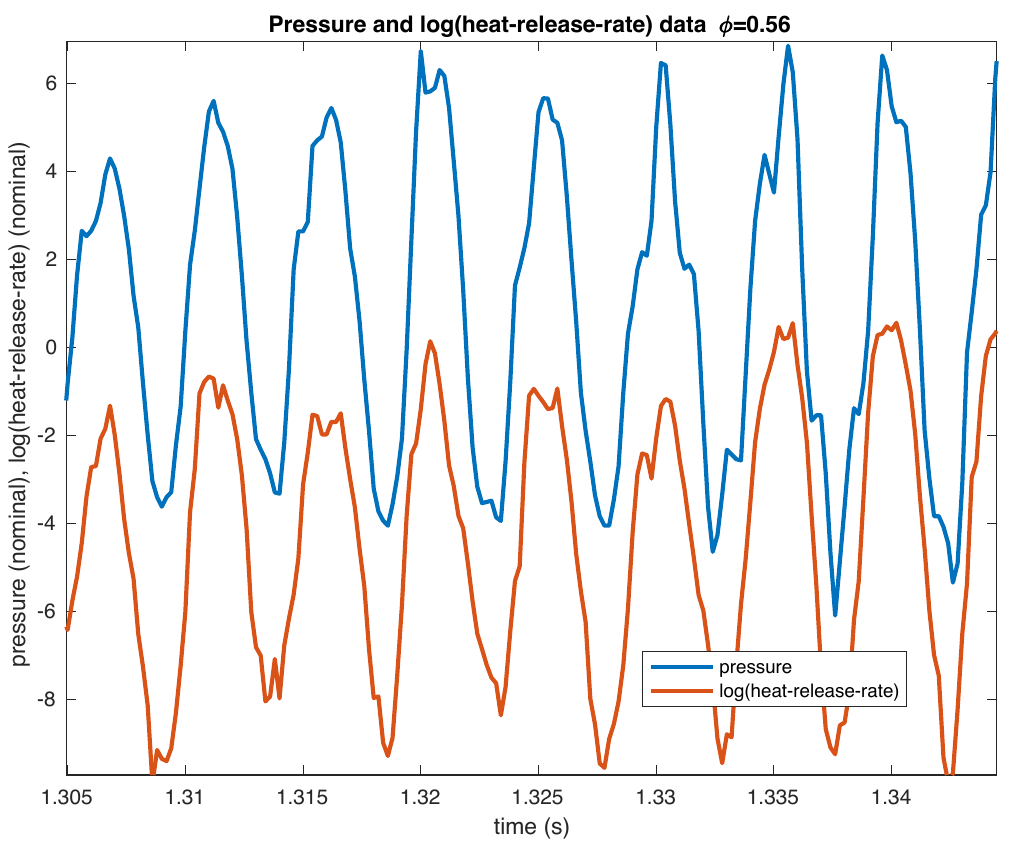}
\caption{\label{fig:dataPlot1}Plot of pressure and log(heat-release-rate) measurements versus time showing sustained oscillatory behavior together with local variability.}
\end{centering}
\end{figure}

The Discrete Fourier Transforms of same data, pressure and log(heat-release-rate), were computed and their magnitude plots, in dB, are shown in Figure~\ref{fig:freqPlot1}.
\begin{figure}[h]
\begin{centering}
\includegraphics[width=0.6\textwidth]{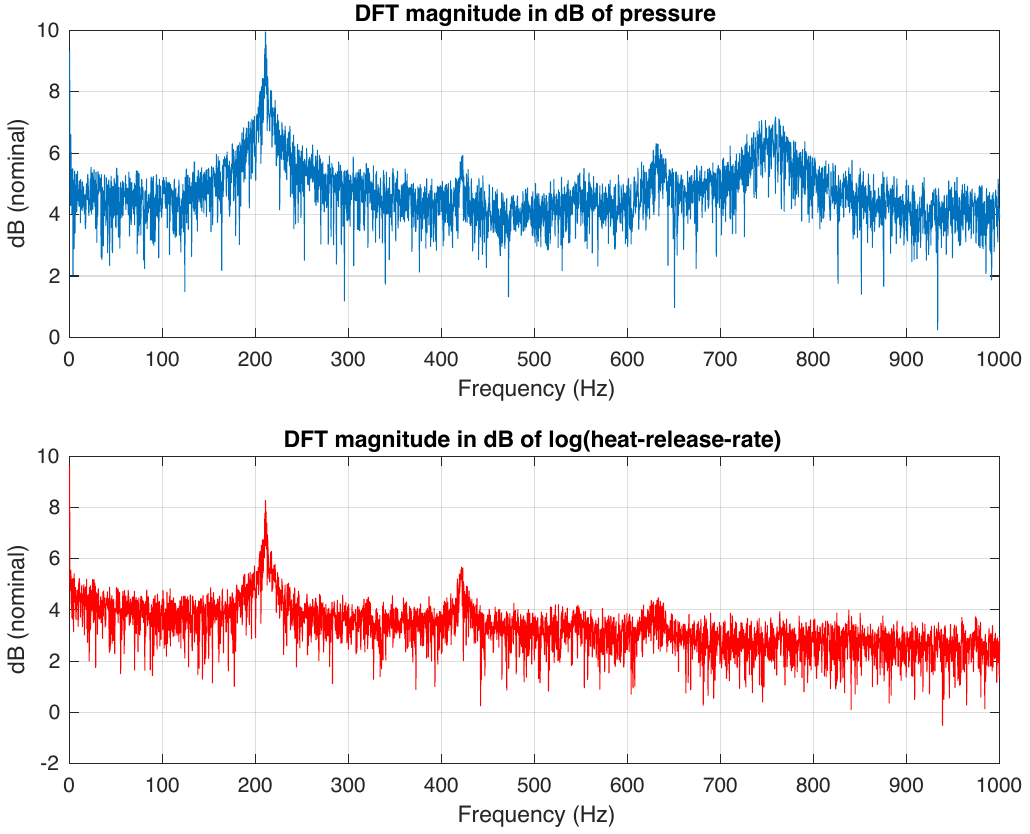}
\caption{\label{fig:freqPlot1}Magnitude plots of the DFTs of pressure and log(heat-release-rate) measurements versus frequency.}
\end{centering}
\end{figure}

From a top-down overview, these experiments served to corroborate the hypothesis that such oscillations occur and that they are affected by equivalence ratio in this way. This information was already widely available \cite{barrereWilliams1969} but here was used to validate the performance of this experimental apparatus. 

\subsection*{Step 2: Inductive analysis of the data}
A number of empirical observations can be made from the experimental data.
\begin{itemize}
\item The data display persistent close-to-periodic elements in pressure and heat-release-rate in both time and frequency domains with evident peaks and spreads in the latter plot.
\item The main signal component is a rough sinusoid around 210 Hz and second, 420 Hz, and third, 630 Hz, harmonics of this frequency are present in plots.
\item A strong, broad but non-harmonic signal around 750 Hz is present in the pressure data but absent from the heat-release-rate data.
\item The absence of the 750 Hz signal from the heat data might be explained by the low-pass nature of the optical sensing instrument.
\item As the equivalence ratio is altered, the frequency components alter but remain present. The following table is from \cite{WD03}. The estimates are derived from the maxima of the DFT magnitudes.
\begin{center}
\begin{tabular}{|c|c|c|}
\hline
$\phi$&\begin{minipage}{18mm}\begin{center}\baselineskip 10pt Harmonic\\frequency\end{center}\end{minipage} Hz&\begin{minipage}{25mm}\begin{center}\baselineskip 10pt Non-harmonic\\frequency\end{center}\end{minipage} Hz\\
\hline
0.56&210.9&753.0\\
0.53&209.8&753.0\\
0.51&201.7&747.5\\
0.49&202.7&744.5\\
0.47&198.4&751.3\\
0.45&179.8&720.9\\
\hline
\end{tabular}
\end{center}
\end{itemize}

\subsection*{Step 3: Deductions from fluid mechanics -- Peracchio-Proscia model }
Computational Fluid Dynamics (CFD) is capable of producing similar predictions of oscillatory behavior but is both costly in computer resources and unreliable in three dimensions. Culick \cite{C-AA-76} and Peracchio and Proscia \cite{peracchioProscia1999} derived, from combustion fundamentals, simplified models capable of generating oscillations. Indeed, the single nozzle rig data are used in \cite{peracchioProscia1999,MJCKJBPP-TechRep-1997}. Based on Galerkin approximation of the fluid dynamics model, they provide the nonlinear, lumped-parameter model illustrated in Figure~\ref{fig:pNp}, without the shaded block, and comprising the feedback interconnection of: a linear forward path of a second-order resonant system, two differentiators and time delay of $\tau$ seconds; with a memoryless nonlinear function $\psi(\cdot)$. The derivation is available in \cite{MJCKJBPP-TechRep-1997}.

\begin{figure}[h]
\begin{centering}
\includegraphics[width=0.4\textwidth]{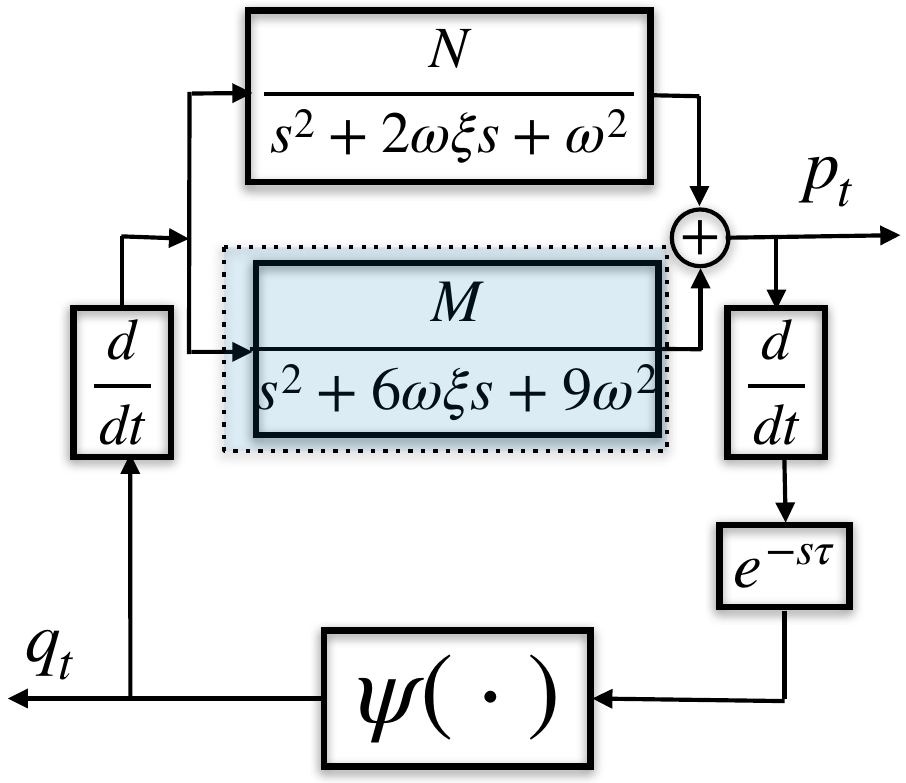}
\caption{\label{fig:pNp}Peracchio \&\ Proscia simplified combustion instability model with pressure signal $p_t$ and heat-release-rate signal $q_t$ \cite{peracchioProscia1999}. The original model did not include the shaded third-harmonic element, which was added in \cite{SBD-IJC-01}.}
\end{centering}
\end{figure}

\subsection*{Step 4: Inductive fitting of the Peracchio-Proscia model}
The Peracchio-Proscia model predates \cite{peracchioProscia1999}, see \cite{MJCKJBPP-TechRep-1997}, and is simply parametrized by $[N,\omega,\xi,\tau]$ plus memoryless function $\psi(\cdot)$. System identification methods were adapted in \cite{MJCKJBPP-ACC-98} to estimate this set of parameters and the function. The data spectrum shown in Figure~\ref{fig:freqPlot1} is just rich enough to fit these four parameter values using the three harmonic frequencies after which the function $\psi(\cdot)$ becomes a graph plot. 

\subsection*{Step 5: Deductive corroboration of the fitted model}
Deductive arguments from the theory of describing functions are brought into play to corroborate the fitted model. The identified linear forward path consisting of the differentiators, resonant system and time delay yield a Nyquist diagram which, when graphed along with the describing function of the identified nonlinearity yields crossing points for the fundamental frequencies of the harmonic components. So, the model survives this test.

\subsection*{Step 6: Refutation and hypothetical replacement of the Peracchio-Proscia model}
While capable of reproducing the first harmonic frequency in the data, the Peracchio-Proscia model is not able to produce both the harmonic frequency and the non-harmonic frequency simultaneously \cite{SBD-IJC-01}. The describing function analysis for the fundamental mode does not extend to capture the non-harmonic signal. Accordingly, a slightly more complex model is proposed, as illustrated in Figure~\ref{fig:pNp} with the incorporation of the shaded block.
The arguments posited for this new hypothesis are based on describing function analysis in \cite{SBD-IJC-01} and amplified in \cite{WD03}. Additionally, the presence of the hypothesized more complex model requires now just one further parameter, $M$, to be identified. Given the presence of exactly three discernible sinusoidal frequencies in the data, these six parameters are at the limit of identifiability of this new structure. This pragmatic limitation is a feature of the new hypothesis; it should be testable with the data.

From our perspective here, the detailed arguments are unimportant compared with the appeal to control theory and postulation to replace one model with a slightly more complicated one. There are two stages: refutation of the earlier model, and then its replacement by a candidate alternative. The new model is an educated guess and not based on the original fluid dynamics. Loosely, the new hypothesis reflects the known behavior of open-organ-pipe acoustics, roughly akin to the combustor geometry, where odd harmonics alone arise.

\subsection*{Step 7: Deductive corroboration via bifurcation analysis}
The presence and persistence of multiple non-harmonic frequencies is noteworthy. Describing function analysis cannot predict the longterm simultaneous persistence of both oscillations. It predicts solely the higher frequency to survive \cite{DB-CDC-03}. More sophisticated tools such as bifurcation analysis were applied to yield Figure~\ref{fig:biff}. The system is described by a nonlinear delay-differential equation and so sophisticated computational tools were required \cite{E-00}.
\begin{figure}[h]
\begin{centering}
\includegraphics[width=0.65\textwidth]{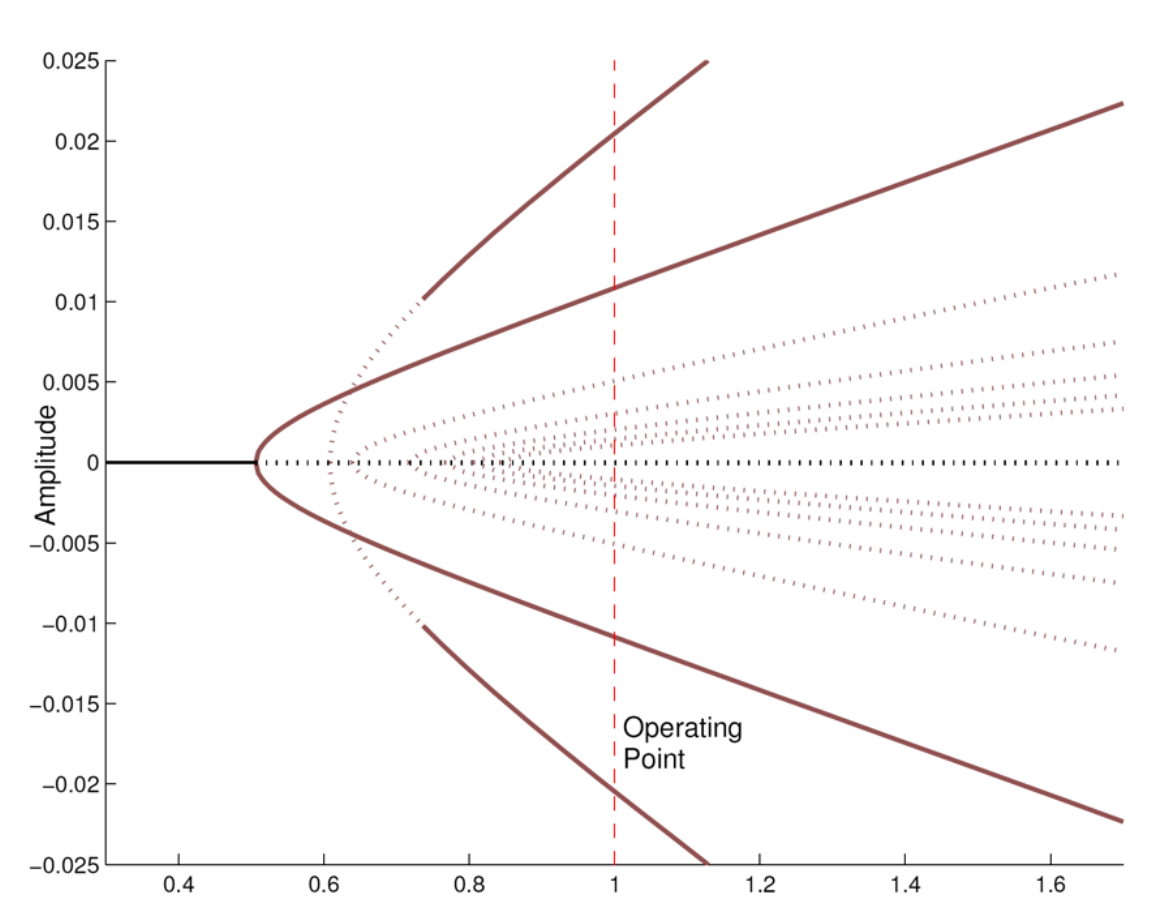}
\caption{\label{fig:biff}Bifurcation diagram of the modified model as a gain inserted in the loop is adjusted \cite{WD03,DB-CDC-03}. At the nominal model ($\mu=1$), both periodic orbits are stable. The diagram show the limits of periodic solutions of one state variable; solid lines mean stable orbits and dotted lines mean unstable. Note, the underlying delay-differential system is infinite-dimensional.}
\end{centering}
\end{figure}

The augmented Peracchio-Proscia model was further developed and analyzed in simulation by including a noise excitation \cite{DunstanBitmeadSavaresiCEP:2001}. The hypothesis was based on the turbulence of the flow in the combustor. So far, these hypotheses are yet to be refuted.

\subsection*{Step 8: Extremum-seeking control -- hypothesis and corroborating experimental test}
The complexity of the as-yet-unfalsified model does not lend itself to control design. Many theoretical and experimental studies in combustion 
\cite{jonesLeeSanatviccaPowerProp1999,annaswamyFleifiletalTCST2000,cohenBanszukJPP2003,dowlingMorgansAnnRevFM2005,balachandranDowlingMastorakosFTC2008} 
point to the capability to diminish oscillations using fuel flow modulation. Banaszuk \textit{et al.} \cite{BAKJ-AUTO-03} hypothesized that modulation of the fuel flow at the fundamental frequency of roughly 210 Hz would suffice for quenching the combustion instability. They devised a control system with the combustion chamber pressure sensor and, as control input, an experimental spinning valve to modulate a part of the fuel flow \cite{barooahAndersonCohenGasTurb2003}. The relative phase of the input to the output at the fundamental frequency was the parameter of this controller. This phase was adjusted slowly using an extremum-seeking approach to construct an estimate of the oscillation amplitude gradient online. The hypothesis was evaluated deductively based on simplified models, although the controller itself is effectively model-free. The results of this test corroborated the approach.

\subsection*{Step 9: Subharmonic and non-harmonic forcing control -- hypothesis, supporting simulation and theory}
The experimental spinning valve operating around 200 Hz would have been difficult to construct for reliable long-term operation in certifiable commercial engines. Following the above reports 
\cite{jonesLeeSanatviccaPowerProp1999,annaswamyFleifiletalTCST2000,cohenBanszukJPP2003,dowlingMorgansAnnRevFM2005,balachandranDowlingMastorakosFTC2008}, non-harmonic forcing around 45 Hz was investigated computationally in \cite{DB-CDC-00} with the modified model. Actuation at these rates would be more feasible commercially, as the valve would operate reliably for longer. Theory was developed in \cite{ioanFethiBobChapter23} to examine the mathematical basis for quenching sustained system oscillations using periodic forcing, with an emphasis on the combustion instability problem. These are deductions flowing from the hypothesized modified model structure. They have yet to be corroborated or refuted in commercially realistic experiments.

\subsection*{Step 10: Passive Helmholtz resonator implementation}
Given an appreciation of the complexity of active control solutions, both in terms of analysis and hardware longevity, passive solutions to redesign of the combustor \cite{gyslingHelmholtz1998} were implemented in products. The role of the exploration into combustion instability characterization and modeling is to provide a sound engineering and commercial basis for making such decisions. Historically, these Helmholtz resonator passive designs preceded the active control investigations. The value of the analysis is that it provides quantification of the design aspects, since the simplified model depicted in Figure~\ref{fig:pNp} contains approximations of the chamber resonant dynamics and time-delay between nozzle exit and the flame front. While the active control was not implemented commercially, it remains available and characterized should commercial or operational requirements change in the future. This aspect of preparedness for future markets is integral to business planning.

\subsection*{Ein Blickwinkel}
In this more comprehensive example the sequence is more exhaustive. We follow the flow of ideas from \cite{WD03,DB-CDC-03,DunstanBitmeadSavaresiCEP:2001} before extending them further with more recent information.
\begin{enumerate}[label=\roman*.]
\item The phenomenon is well known in laboratory scale experiments and evident in the single-nozzle rig meant to emulate combustion in commercial engines. The physics and fluid dynamics of flow and combustion again rely on classical formulations \cite{WilliamsCombustionTheory1994} yielding partial differential equations and, ultimately, computational fluid dynamics models, where the phenomenon is again present, but difficult to compute reliably. This is the epistemological starting point.
\item A set of experiments was conducted with differing equivalence ratios. Interpretation of this data yielded the inductive analysis of Step~2.
\item The subsequent steps comprise a sequence of hypotheses, deductions, experiments, inductive fits to the data or refutations.
\item The control design included the extremum seeking non-model-based approach hypothesized on the gradient calculation. This was extended to subharmonic forcing based on the (still highly simplified) extension to the Peracchio-Proscia model and the deductive methods of \cite{ioanFethiBobChapter23} for quenching. In turn, this was compared with passive methods and evaluated with a business perspective, including maintainability and certification.
\end{enumerate}

In comparison to the servomotor control problem, this example lies closer to the scientific method. Actuation and sensing limits dictate that simple models be sought, but the sequence of steps sees replacement of earlier models and increasing degree of sophistication in the model complexity and in the analytical tools required for its assessment. The substitution of the Peracchio-Proscia model by its so-far-unfalsified extension would hardly count as a scientific revolution according to Thomas Kuhn. And the objective is respectfully limited to focus on capturing the dominant phenomena rather than on predictive properties. This reflects the eventual controller emphasis. It is worth noting too that the model complexity is kept very low, less than that suggested by Rissanen's Minimum Description Length criterion, to manage controller gains resulting from the design tools.

The several sets of experimental data are carefully designed and evaluated with tools of narrow scope, given the limitations of the heat-release-rate sensor and the eventual controllers. The data were expensive to acquire and, therefore, used repeatedly in inductive testing as the hypotheses were refined. The interplay between the scientific phases is intricate and very far from purely inductive.
}

{
\section*{Sidebar: Control of a sugar mill}\label{sidebar-sugarmill}
The system identification and model-based control design of a sugar mill at CSR (now Wilmar Sugar) Victoria Mill in Ingham Queensland Australia is described in detail in \cite{PAR1995,Bitmead2002}. Figure~\ref{fig:macknade} shows the cascaded five-mill milling train from Wilmar Sugar's Macknade Mill, about 10 km northeast of the larger Victoria Mill. 

\begin{figure}[h]
\begin{centering}
\includegraphics[width=0.8\textwidth]{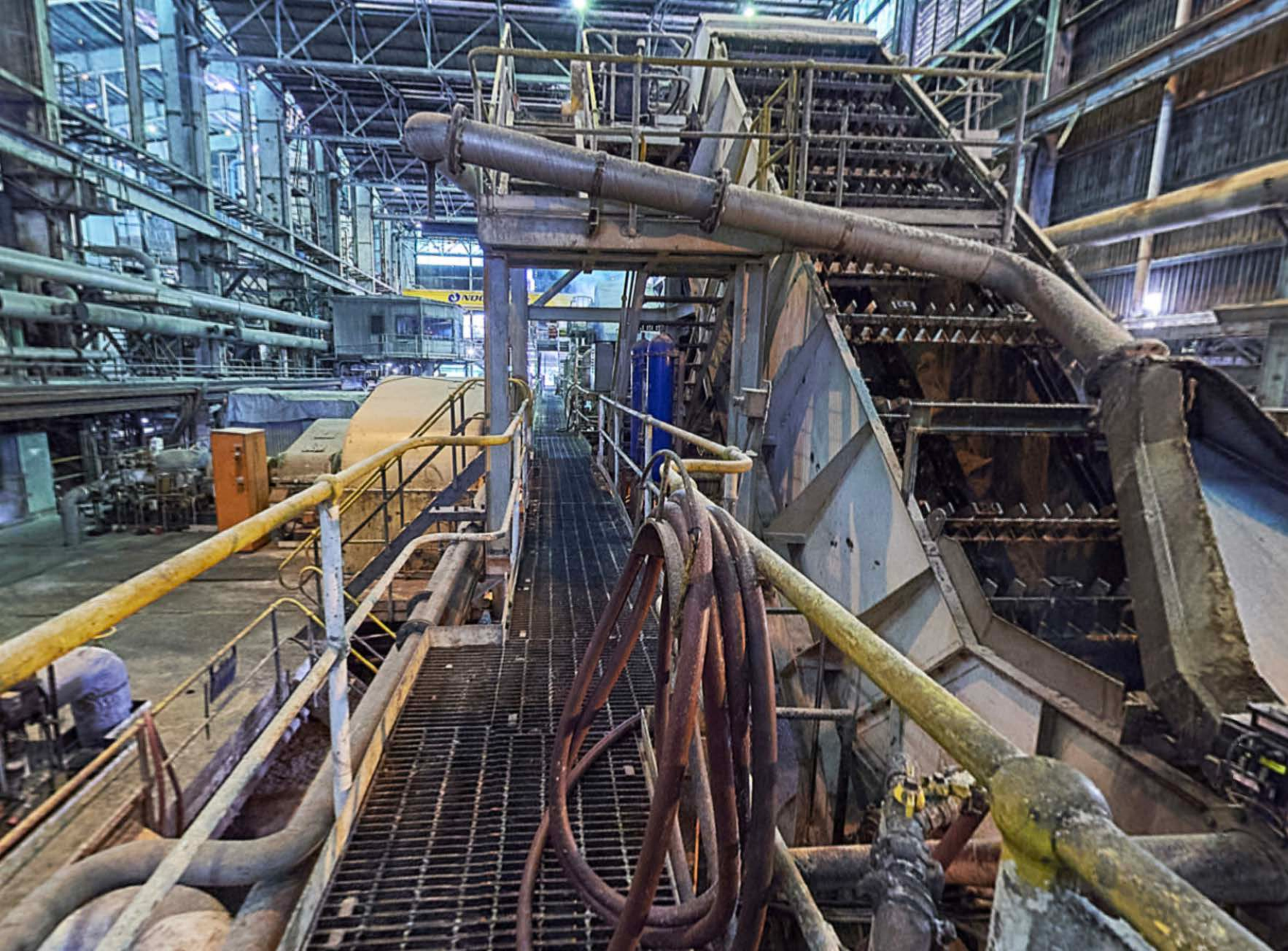}
\caption{\label{fig:macknade}Wilmar Sugar's Macknade milling train, from \href{https://www.wilmarsugar-anz.com/360}{https://www.wilmarsugar-anz.com/360}. This milling train comprises five mills in series each of which is a two-input/two-output system. The foreground shows the elevator which delivers cane from the preceding mill or shredder into the top of the feed chute of the mill.}
\end{centering}
\end{figure}

The objective of the mill is to extract the juice from sugar cane by crushing many thousands of tonnes of cane per day throughout the June-December crushing season. Each mill is controlled by two inputs, delivery chute exit gap managed by a hydraulic flap and mill speed signal for the drive motors. The measured signals are the level of material in the chute, determined by conductivity probes, and the torque of the mill, computed from speed and steam pressure. The quantified objective is to maximize the mill torque within an upper constraint while maintaining the chute level within upper and lower limits. The control problem is dominated by the large variability in the mechanical properties of the differing varieties of sugar cane grown by the farmers, with their own objective of maximizing financial returns. As the varieties vary, the slippage of the mill rollers on the mat of soaked cane fiber changes, affecting the extraction of sugar juice, mass throughput and the capability to support high torque.

Here are the pertinent features of the iterative design.
\begin{description}
\item[Starting point:]
\begin{itemize}
\item The sugar mill is available for experiments but is open-loop marginally unstable due to the presence of integrators in the milling process; basically, the chute acts as a small tank. The preexisting decoupled controller consisted of a PI feedback between chute height and speed and a PI controller between torque and flap position. The integrators in plant model and controllers indicate the step-like nature of the disturbances. The sampling rate is 1 Hz. The experiment requires the presence of cane being crushed to yield useful data. That is, the system cannot be tested in a disturbance-free environment.
\item The initial closed-loop is dominated by the disturbances leading to excursions hitting the chute-full and chute-empty constraints. These cane-variety-dependent disturbances persist for between 5 and 15 minutes. 
\item The control design objective is to achieve high juice extraction rates, which are linked to maintaining high mill torques. So, minimum variance strategies are sought to minimize constraint violation, notably in the chute height, when the torque set point is high. Statistical evaluation of the controller performance requires closed-loop data from over twelve hours in order that the full distribution of disturbances is seen.
\item The existing controller was regarded as too slow -- bandwidth less than 0.05 Hz -- in response to changes in cane type, leading to large swings in the output variables. A new controller designed using both inputs based on both outputs was sought using a linear identified model and LQG MIMO feedback controller design. 
\end{itemize}
\item[Excitation signals:] Because the feedback control aim is to extend the bandwidth of the closed loop, excitation signals in both channels are designed to be high-bandwidth with significant energy up to 0.2 Hz. To yield visibility in the measured outputs in the face of disturbances, these excitations themselves need to be significant in range. Both excitations were filtered white noise signals with speed excitation being 30\% of full scale and flap excitation 20\%.
\item[Initial identification experiment:] With both PI controllers in place, the system is perturbed by adding the excitation signals to the feedback signals to generate closed-loop data.
\item[First identified model $\hat P_0$:]  Using data down-selection methods and filtering described in \cite{Bitmead2002}, a Box-Jenkins model is fitted using careful model structure selection. The plant model, $\hat P_0$, is kept and the fitted disturbance model, used to capture signals outside the input-output description, replaced by the known disturbance characterization based on deductions (thought experiments) on the target system and supported by the extant PI controllers.
\item[First control design $C_0$:] An LQG controller, $C_0(\hat P_0)$,  is designed using scaling to balance between the separate signal channels and disturbance models incorporated into the model. 
\item[Closed-loop data analysis:] The LQG controller $C_0$ is inserted in place of its PI predecessor and data collected in closed loop without signal excitation. The ensuing control performance in closed-loop is disappointingly poor.  However, the model together with the LQG design yield not just the controller but also an expectation of the closed-loop signal spectra. Comparison between the achieved closed-loop spectra and their predicted or designed values indicates that the plant model fits poorly outside the very low frequencies of the PI controller. Using such a model for higher bandwidth control is ineffective. Figure~\ref{fig:cloopspeed} depicts the power spectral densities of the closed-loop speed signal with the PI controller and with LQG controller $C_0$. Note that these closed-loop data are unexcited but are driven by the cane-variety disturbance process.
\begin{figure}[h]
\begin{centering}
\includegraphics[width=0.5\textwidth]{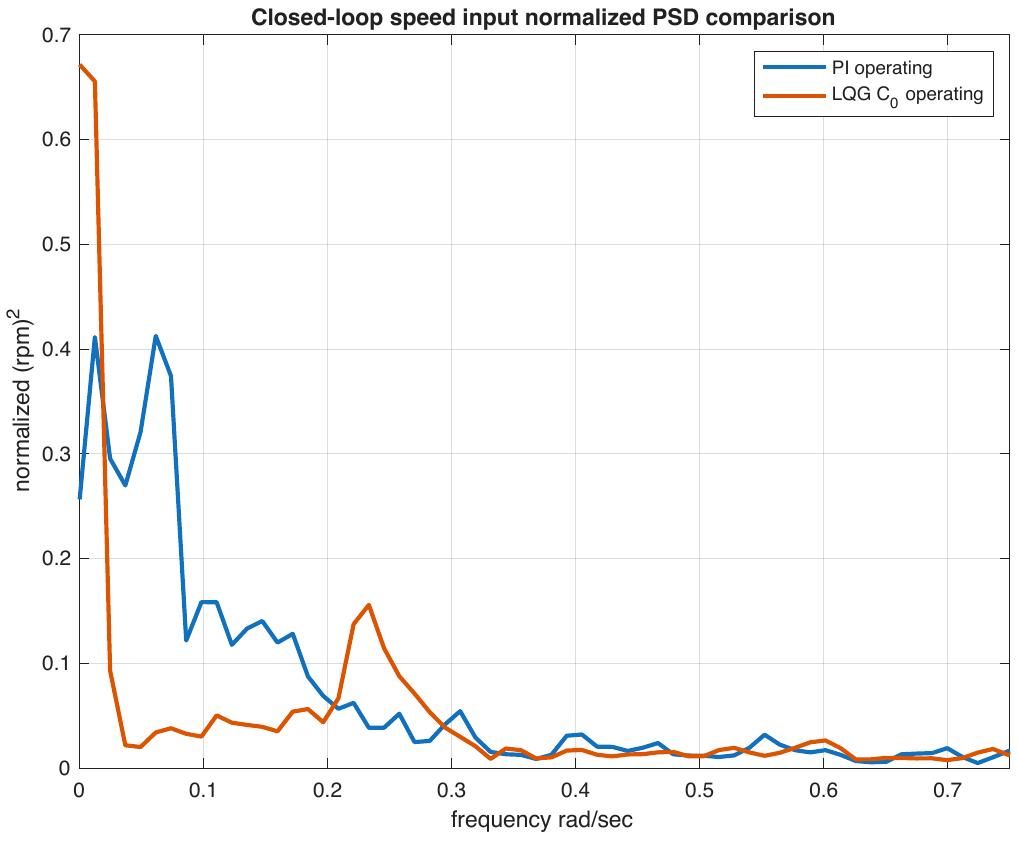}
\caption{\label{fig:cloopspeed}Closed-loop speed signal power spectral density comparison with PI-control and LQG controller $C_0$. The attempted-higher-performance controller $C_0$ alters the closed-loop signal spectra, which in turn focuses subsequent model fitting in different frequency ranges; here, closer to the designed gain crossover frequency of 0.25 radians per second.}
\end{centering}
\end{figure}

\item[Second control design $C_1$:] Using model $\hat P_0$ and frequency weights derived from the designed and achieved closed-loop performances from the preceding closed-lop data, a new frequency-weighted LQG controller, $C_1(\hat P_0)$ is constructed and rejected after trial.
\item[Second identification experiment and model $\hat P_1$:] Next with the LQG controller $C_1$ in place in the feedback loop and the perturbing excitation signals from earlier, a new closed-loop experiment is conducted. Controller $C_1$, since it was designed to extend the bandwidth, yields more informative data for a new model, $\hat P_1$, to be fitted from which a new controller might be designed. This model uses the new closed-loop experimental data and applies the indirect method of system identification \cite{VS-A-93} again with a Box-Jenkins model structure, as before.
\item[Third controller design $C_2(\hat P_1)$:] A non-frequency-weighted MIMO LQG controller was designed and tested. This is the final controller, which was implemented and accepted.
\item[Coda:] Controller $C_2$ performed well in trials and was able to reject the disturbance effects rapidly. A fortuitous by-product of the design was significantly reduced energy consumption by the mill. At the time, this was not a major factor in assessment of the controller. However, since the energy consumed by the mill is produced by burning the residual fiber to produce steam to run the factory and electricity, which is exported to the grid, this electricity is a bio-energy which is increasing in value.
\end{description}

\subsection*{Ein Blickwinkel}
This large-scale control application led to many ideas behind iterative identification and control design \cite{AlbertosSala:02}. 
\begin{enumerate}[label=\roman*.]
\item The sugar mill is a longstanding industrial process which involves many mechanical phenomena associated with extraction of juice by breaking sugar cane cells. There is no adequate first-principles model available. Although, the two-channel PI controller does function but poorly. Disturbance signals are significant and relate to the varying mechanical properties of the sugar cane.
\item Experiment design is a central feature and is based on quite disruptive excitation signals to subvert at times the otherwise dominant disturbances. Still, the data needs to be carefully curated and downselected. All data are not equally informative for the purpose.
\item The feedback controller in operation during closed-loop experiments affects the data content and the nature of the model fit. The underlying nonlinearity of the target system coupled with fitting linear models mean that the effect of the controller on data content is not remediable by filtering.
\item The sequence of steps is not reflective of improving solely the model's predictive capabilities but of adjusting a fixed-complexity model to manage modeling error for the current feedback controller.
\item Both the identified approximate model and its controller are adapted iteratively. The design terminates after a few steps.
\end{enumerate}

This example stresses the distinction between science-oriented modeling, albeit with a view to preserving manageable complexity, and the contention that the successive models should have a quality of fit adapted to the needs of the controller. In turn, these needs are dictated by the controller itself. So a solution is necessarily iterative. The informativity of the data for the control purpose changes with the feedback controller in operation during the experiment. This is evident in Figure~\ref{fig:cloopspeed} with operational, i.e. unexcited, closed-loop data with different controllers.

The data collection is highly non-routine and selective, being affected by the controller in place, the excitation signals, and the prevailing disturbances. It is not particularly expensive to collect, although some diminution in extraction performance during the experiments is expected but not quantified. Post-experiment, a significant quantity of data is removed following inspection and assessment. Older data sets are removed from further consideration in their entirety once a more highly performant controller is devised. In this application, the curation of the data is central to the success of the control. Certainly, there is no equivalence between the several data sets, with some being totally deprecated or defunct. Successive models and controllers replace totally earlier models and controllers. There is no subsumption or nesting of models. This conforms to Kuhn's scientific revolution \cite{kuhnBook1962} at every stage.
}

{\section*{Sidebar: Countermanding experience}
\begin{figure}[h]
\begin{centering}
\includegraphics[width=0.5\textwidth]{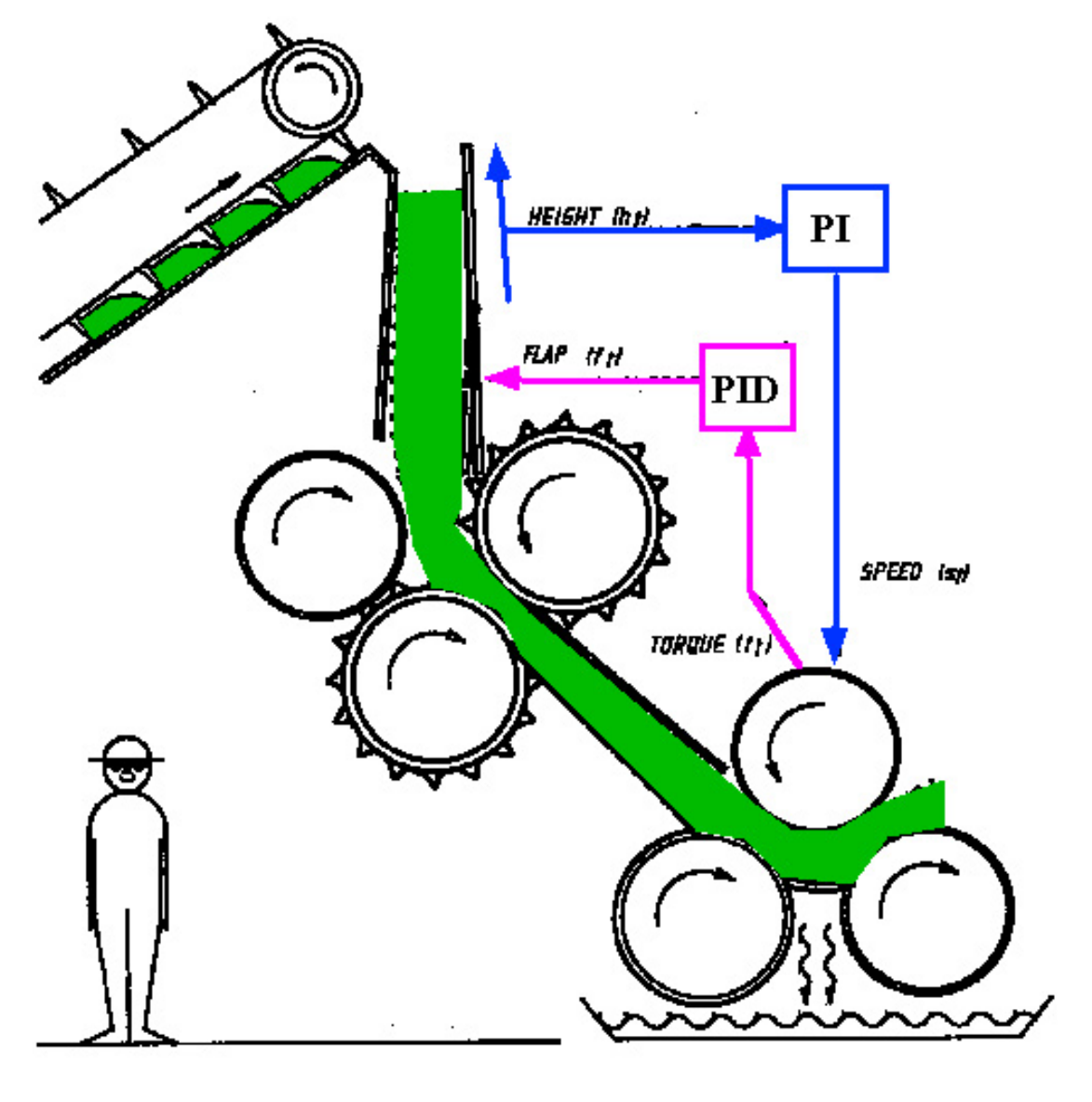}
\caption{\label{fig:millman}Side elevation diagram of a sugar mill, with human for scale, indicating the input (speed and flap) and output (chute height and torque) variables and their original interconnection via two single-input-single-output controllers.}
\end{centering}
\end{figure}

Figure~\ref{fig:millman} depicts a single sugar mill, from a cascaded sequence of five at CSR Victoria Mill in the mid-1990s. It shows the flow of wetted and shredded sugar cane moving from upper-left to lower-right via: the elevator, chute, flap aperture, toothed power-feed rollers, the constrained channel, the three-roller mill itself, and then out either to the soaking `boot' and subsequent mill or to the boiler for incineration and steam production. Figure~\ref{fig:macknade} shows a similar current mill looking at the elevator. Figure~\ref{fig:millman} also shows the starting point for the feedback controller; two single-input-single-output PI controllers, one between chute height and speed and the other between torque and flap opening. This legacy controller provided the starting point for the two-input-two-output controller design in \nameref{sidebar-sugarmill}.

Physical arguments and engineering experience underpin the choice of this legacy control architecture.
\begin{description}
\item[Height-speed loop]: When the chute height exceeds its nominal value, speeding up the mill will increase the material flow and regulate the height back to set point.
\item[Torque-flap loop]: When the mill torque exceeds its nominal value, choking the flow of material via the flap should reduce the torque towards set point.
\end{description}
This control formulation functioned adequately for many years and became progressively unsatisfactory as the throughput of the mill was increased; the response time seemed too slow to accommodate cane variety changes to material mechanical properties, hence the impetus for control design discussed in \nameref{sidebar-sugarmill}. At this stage, it is pertinent to reiterate the presence of integrators in the two control loops, which is evidence of the presence of step functions in the disturbance process. That is, cane variety changes and their associated mechanical feeding occur infrequently, say over multiples of five minutes, and as step changes. This characterization of the disturbance was used in the subsequent control design.

However, successive system identification closed-loop experiments with excitation followed by model fitting repeatedly yielded models with the following roughly cross-diagonal frequency-domain structure ($\epsilon$ indicates a small quantity) for the plant system
\begin{align*}
\begin{bmatrix}H(z)\\T(z)\end{bmatrix}&=\begin{bmatrix}\epsilon X_{hs}(z)&X_{hf}(z)\\X_{ts}(z)&0\end{bmatrix}\begin{bmatrix}S(z)\\F(z)\end{bmatrix}+\begin{bmatrix}D_h(z)\\D_t(z)\end{bmatrix}.
\end{align*}
The legacy control structure would be suggested by the transfer function matrix above being effectively diagonal. Experience and mental models conflict with the inductive fit.

\subsection*{Experiment design and inductive evaluation}
The hypothesis to be tested is whether, indeed, the open-loop plant transfer function from flap to torque is effectively zero. While repeated closed-loop experiments indicate this, a more incontrovertible experiment was required to convince the engineers of this feature. Accordingly, an open-loop step test was performed. The instability of the plant lies in the presence of integrators in subsystems and in the disturbance process. It is possible to operate the mill in open loop for minutes at a time without chute height excursions reaching saturation.

The mill was operated under feedback control until it reached stable operation around the lower limit of chute height. Then feedback was removed and a step applied in the flap variable. The mill torque did not alter appreciably in response to the step, although the chute height ramped upwards with the closing flap. Modest disturbances still were in effect during the experiment and torque did drift up and down. After about eight minutes, the feedback was reengaged when the chute height reached close to its upper limit. 

The experiment confirmed the zero flap-torque transfer function refuting the non-zero transfer function hypothesis and corroborating the zero value. It also, corroborated the presence of an integrator in the flap-chute path.

\subsection*{A new candidate view}
Following the open-loop experiment and building on the models identified in closed-loop, a new physical explanation was formulated.
\begin{description}
\item[Chute height:] is strongly dependent on flap position, since the cane arriving provides a fixed mass-flow in and the flap determines the mass-flow out. An integrator is therefore present. There is a weak dependence of the chute height on speed inasmuch as it affects the flow of cane from the bottom of the chute; the friable nature of the cane mat suggests that there is little pull from the mill. The signs are as imagined.
\item[Torque:] is primarily dependent on the speed of the mill and almost independent from the flap position. This is a complicated relationship incorporating both acceleration and slippage effects dependent on cane variety.
\end{description}

How should one understand the functioning of the old two-PI controller function? A Gedankenexperiment helps. Suppose, from steady operation, a step change in cane variety occurs causing the chute height to rise; that is, mass-flow from the chute exit is diminished, since the inflow is constant. Then, the chute-speed PI controller causes an increase in speed, which in turn causes a diminution of torque leading, via the torque-flap PI controller, to further opening the flap, thereby arresting the increasing chute height. The control action has passed through both PI controllers and the speed-torque plant subsystem. The result is a much delayed but correct control action responding to a step in chute height by opening the flap. So the replacement viewpoint is consistent with past performance and holds out the prospect of much faster controller response.

\subsection*{Ein Blickwinkel}
Experience informs engineering design and, in the current framework, much of the hypothesis setting. The history of science provides many examples of where experience was misleading and led to replacement of theories -- the physicists are fond of calling them \textit{laws}. The example here is included to indicate the presence of the same dynamic in control and its replacement, perhaps only temporarily, by a different hypothesis following an exacting experiment. A salient feature is that, even with the promotion of a new and better supported viewpoint, not everyone is so easily convinced. From a control perspective, the identification of a low-delay control path led to improved performance. But human factors require work to alter the paradigm.
}

\section{Industrial examples and the role of data in control}\label{section:Data_in_Control}

To add some meat to the bones of the philosophy of control, four commercially significant practical examples of sensor and control design, both model-free and based on modeling, are explored, the objective being to examine the scientific and engineering processes and their explicit data requirements rather than the specifics of the problems. Each sidebar terminates in a Blickwinkel (or \textit{perspective} in German) highlighting aspects of the data and their role in the control design process.

Evidently, feedback control relies on data to feed back. Although, parametric or vibrational control, which might include the non-harmonic forcing in the combustion instability example is unlikely to be considered feedback. The \textit{model-based} approach would use experimental data to fit a deductively chosen model structure using the methods of system identification, as is the normative (\textit{c.f.} the DC servomotor), but perhaps practically uncommon, paradigm. These data would need to be informative for that structure, as is quantified by the Fisher Information Matrix or an empirical counterpart, and the model itself would need to be presumed to extrapolate to the operating data environment of the next controller, such as avails when the target system is essentially linear. In this circumstance, many data sets are more or less equivalent and suffice for fitting models and then designing controllers.

In system identification, one usually conceives of two independent data sets from the same stationary distribution for input and disturbance signals; an evaluation set and a validation set. The model is fitted to the evaluation set and then its predictive powers are assessed on the validation set. A similar approach is adopted in much of machine learning, although there are significant departures in views of overfitting the data with very complex models. A problem arises when a change of controller causes the closed-loop data properties to alter significantly, as was seen with the sugar mill. Now, data from past experiments from a different operating regime are no longer relevant for modeling or subsequent control design. Effectively, we have lost the capacity to extrapolate the prior models' behaviors to the new environment. Thus, new data are required.

A message from the sugar mill example, and reflected in the closed-loop signal spectra, is that data need to be informative for their purpose and accommodate the ultimate utility. In this regard, not all data are equivalent. For the sugar mill, the data collected in closed loop with the initial PI controllers in place were poorly suited to fitting a model which was accurate around the eventual crossover frequency. New experiments, depicted in Figure~\ref{fig:cloopspeed}, under a higher bandwidth controller were needed to adapt the quality of fit to the controller requirements. 

Considering the extremum-seeking combustion controller, the data information content needs to suffice for computation only of the local gradient of performance with respect to the control parameter. Hypotheses need to be made about the local gradient and its observability via the experiment but the successive experiments are conducted at differing operating points, i.e. phases, of the controller.

Obtaining informative data comes at a cost to control performance during the experiments. For continual adaptation of models, this appears as the persistence of excitation condition familiar from many years of adaptive control theory. When adaptive control theory becomes adaptive control practice, as in mobile and agile wireless systems, a pragmatic method to ensure continual excitation is to include a non-information-bearing \textit{midamble} sequence into every data packet \cite{KruegerDenkYangMidamble2006,LeeBahkMidamble2021}, which comes at a serious cost to the data capacity but without which functioning ceases. 

In many papers on data-driven control, this persistence of excitation condition is stated as a rank condition and is associated with Willem's Fundamental Lemma \cite{WillemsRapisardaMarkovskyDeMoorSCL:2005,MarkovskyWillemsVanHuffelDeMoor:2006}, which in turn deals solely with exact linear system identification, i.e. noise-free data, zero disturbances and exact modeling. It has been extended to a more quantitative form in \cite{berberichEtAlPEACC2023}, where it resembles persistence of excitation conditions. From the examples, we see that there are other underlying aspects, such as linearity, known finite dimension, absence of noise, which form part of the epistemological background to these methods. For the inductively powered commercial product FrothSense+, the historical training data are representative of the stationary operation of floatation separators including their variability with disturbances when running; yet statistical post-validation remains important.

In more general terms of the data-dependent world of artificial intelligence, machine learning and big data, the message is much more conditional. Data come in different flavors and utilities for purpose. Static or statistically stationary environments are needed in order for data sets to maintain informativity across distinct sets or applications. This can, of course, be the case but needs to be hypothesized and, perhaps, refuted. As the model classes reach a level of complexity that they might capture many facets of target system behavior, still their utility for purpose needs to be evaluated. Some data becomes defunct as the operating environment changes, which is not to suggest that it remains benign if it remains in play. Tying the quality and validity of data and its purpose in prediction and control is a central aspect of data curation, generation and (experiment) design.

\section{Joining the threads}
Control engineering is a practical pursuit with much in common with its cousin empirical science. So, the history and philosophy of science provides a useful framework to consider the role of experimental data within this methodology. Science provides serious cautions, but certainly no prohibitions, concerning purely inductive or data-driven methods. Central among these is the uniformity principle of Hume and its presumption that (past-)data-driven methods necessarily embody the complete and immutable totality of the system and its future operation. That is quite an assumption and evidently untrue in general for routine operational process data. Willem's Lemma tells us that. But it also rests on other stringent and often unstated assumptions. Since lemmata belong to the deductive world of theory, there is a need for care in unchecked reliance on these assumptions in practice. The examples of the sidebars are there to indicate the practical side of applying theory and experimentally derived data in control design. The take-home message is that high quality data suited to control design is frequently difficult to collect and hard to recognize. Even more, its value for control design is context and objective dependent, which can attenuate its worth over time.

With the hucksters, shysters, touts and carnival barkers for `artificial intelligence' rife in the press and literature, there is need for care and reflection in adopting purely inductive approaches to feedback control design. `Big' is not a useful quality of data to capture its suitability for purpose in control. Melanie Mitchell, an AI insider, in her recent book \cite{mitchellAI2019} provides a cautious historically informed but forward looking evaluation of AI performance and limitations, particularly in the realm of what success looks like in this area; say 85\%-correct image recognition for cats or stop signs. Does such a strike rate work for control? It depends on the problem. Are more data necessarily better? That depends too \cite{CarrettBastinGeninGevers_TSP:1996}.

Metso Minerals' FrothSense+ is described in a sidebar. It is a commercially valuable instrument in assessing the performance of floatation cells and thereby controlling them. The success of this sensor is quantified via correlation analysis with specific (never exceeding 0.9) coefficients of determination, $R^2,$ values \cite{visioFrothSlides2007}. Machine learning was used to train the sensor on laboratory validated data and it operates well in a safety-non-critical environment. Although, it is not involved in the control law determination. So, inductively assisted controllers can function to great benefit where experience dictates that Hume's concerns are not germane; that is, the process and disturbances are stationary and the data are representative. The statistical performance is appropriate to the process objective of ore recovery over multiple stages.

The DC servomotor example also provides reinforcement for data-driven methods and Willems Lemma. The limits to operation, low-noise signals, system complexity and linearity are all well appreciated. As is said in the sidebar, this is the archetype conveyed for how control design should function. Provided future operation conforms to these limits in training, the controller should and does work.  It is informative to read \cite{MarkovskyHuangDorflerCSM2023,CarletBolognaniDorflerIEEETPpowerElect2022} of applications of data-driven methods and their successful implementation, after deliberate disruptive experiments during data collection, in environments with well appreciated closed-loop dynamics and their complexity. According to our restricted usage of the term \textit{data-driven}, these works describe much more than purely inductive methods. They indicate the value of marriage of data-driven methods with appropriate problem domain knowledge. 

When more demanding (of the modeling) processes arise, as in the other two industrial examples, and particularly where approximation and estimation are involved, experiment design and model structure determination bend strongly towards the scientific path mixing experience, hypotheses, deduction, experiment design and induction. The process is almost never one-shot for the reasons detailed in the text above. Data generation and selection becomes a central piece of the design puzzle as the objective shifts to hypothesis refutation. Even in the most rudimentary control applications, validation of performance and commissioning of the controller is a key, but additional, inductive aspect of the engineering task.

The examples in this paper were chosen to reflect the commercial and practical reality of control systems, where reliability, cost effectiveness and commissioning limitations drive decisions and time frames. Such considerations put feedback control at a remove from science. The examples also highlight the combination of design, experience, deduction, experiment, induction and art that makes control systems challenging and rewarding. Given that the majority of important control applications need to accommodate uncertainty in both the plant system and in the disturbance environment -- see the section Intentions of Control near the beginning -- one might expect good future job security for control engineers well versed in statistical methods of modeling, estimation and control, hype notwithstanding.

Given this paper's emphasis on the supra-algorithmic nature of control design and its customary dependence on experience and intricate data generation and curation, it is appropriate to bookend the paper with another quote from Karl \AA str\"{o}m \cite{karlCarpentry} delivered in response to a question regarding experiment design, data down-selection and filtering in modeling for control: ``Why can't I find all this in a book?''
\begin{quote}
\textit{``... because system identification is like carpentry; it cannot be learnt from a book.''}
\end{quote}

\section*{Acknowledgements}
Part of this work was performed at the Institute for System Theory and Automatic Control at the University of Stuttgart whose hospitality and support are valued. Funding from the Alexander von Humboldt Foundation and from the Cluster of Excellence EXC 2075 ``Data-Integrated Simulation Science'' is gratefully recognized. \raisebox{-1.7mm}{\includegraphics[width=10mm]{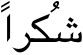}} to Hosam Fathy from University of Maryland for providing the attribution for the opening quote from Karl \AA str\"om. The closing quote occurred during the author's own presentation. The author is indebted, and not just grateful, to his colleagues who provided critical insights and suggestions for improvement; they know who they are! And so should you: Victor Solo, Gonzalo Rey, Rick Johnson, Richard Murray, Ari Partanen, Pramod Khargonekar, Andrea Iannelli, Robert Kosut, Sam Crisafulli, Wayne Dunstan, Allan Connolly, Max Crisafulli.

\bibliographystyle{IEEEtran}
\bibliography{/Users/bob/tex/bobilby}

\end{document}